\documentclass[twocolumn]{aastex631}
\usepackage{amsmath}
\usepackage{rotating}

\newcommand{\um}{\,\micron}
\newcommand{\oIII}{[O\,\textsc{iii}]}
\newcommand{\sII}{[S\,\textsc{ii}]}
\newcommand{\sIII}{[S\,\textsc{iii}]}
\newcommand{\neII}{[Ne\,\textsc{ii}]}
\newcommand{\neIII}{[Ne\,\textsc{iii}]}
\newcommand{\nIII}{[N\,\textsc{iii}]}
\newcommand{\feII}{[Fe\,\textsc{ii}]}
\newcommand{\HII}{\ion{H}{2}}

\shorttitle{FIR Metallicity in M101}
\shortauthors{Lamarche et al.}

\begin{document}

\title{Direct Far-Infrared Metal Abundances (FIRA) I: M101}

\correspondingauthor{Cody Lamarche}
\email{cody.lamarche@utoledo.edu}

\author{C. Lamarche}
\affil{Ritter Astrophysical Research Center, Department of Physics and Astronomy, University of Toledo, 2801 West Bancroft Street, Toledo, OH 43606, USA.}

\author{J. D. Smith}
\affil{Ritter Astrophysical Research Center, Department of Physics and Astronomy, University of Toledo, 2801 West Bancroft Street, Toledo, OH 43606, USA.}

\author{K. Kreckel}
\affil{Astronomisches Rechen-Institut, Zentrum für Astronomie der Universität Heidelberg, Mönchhofstraße 12-14, 69120 Heidelberg, Germany.}

\author{S. T. Linden}
\affil{Department of Astronomy, University of Massachusetts at Amherst, Amherst, MA 01003, USA.}

\author{N. S. J. Rogers}
\affil{Minnesota Institute for Astrophysics, University of Minnesota, 116 Church St. SE, Minneapolis, MN 55455, USA.}

\author{E. Skillman}
\affil{Minnesota Institute for Astrophysics, University of Minnesota, 116 Church St. SE, Minneapolis, MN 55455, USA.}

\author{D. Berg}
\affil{Department of Astronomy, University of Texas at Austin, 2515 Speedway, Austin, TX 78712, USA.}

\author{E. Murphy}
\affil{National Radio Astronomy Observatory, Charlottesville, VA 22903, USA}

\author{R. Pogge}
\affil{Department of Astronomy, The Ohio State University, 140 W 18th Ave., Columbus, OH 43210, USA.}

\author{G. P. Donnelly}
\affil{Department of Physics and Astronomy, University of Toledo, 2801 West Bancroft Street, Toledo, OH 43606, USA.}

\author{R. Kennicutt Jr.}
\affil{Steward Observatory, University of Arizona, Tucson, AZ 85721-0065, USA.}
\affil{George P. and Cynthia W. Mitchell Institute for Fundamental Physics \& Astronomy, Texas A\&M University, College Station, TX 77843-4242, USA.}

\author{A. Bolatto}
\affil{Department of Astronomy, University of Maryland, College Park, MD, USA.}

\author{K. Croxall}
\affil{Expeed Software, 100 W Old Wilson Bridge Rd, Suite 216, Worthington, OH 43085, USA.}

\author{B. Groves}
\affil{Research School of Astronomy and Astrophysics, Australian National University, Canberra, ACT 2611, Australia.}
\affil{International Centre for Radio Astronomy Research, University of Western Australia, 7 Fairway, Crawley, 6009, WA, Australia.}

\author{C. Ferkinhoff}
\affil{Department of Physics, Winona State University, Winona, MN 55987, USA.}

\begin{abstract}

Accurately determining gas-phase metal-abundances within galaxies is critical as metals strongly affect the physics of the interstellar medium (ISM). To date, the vast majority of widely-used gas-phase abundance-indicators rely on emission from bright optical-lines, whose emissivities are highly sensitive to the electron temperature. Alternatively, direct-abundance methods exist that measure the temperature of the emitting gas directly, though these methods usually require challenging observations of highly-excited auroral lines. Low-lying far-infrared (FIR) fine-structure lines are largely insensitive to electron temperature and thus provide an attractive alternative to optically-derived abundances. Here, we introduce the far-infrared abundances (FIRA) project, which employs these FIR transitions, together with both radio free-free emission and hydrogen recombination-lines, to derive direct, absolute gas-phase oxygen-abundances. Our first target is M101, a nearby spiral-galaxy with a relatively steep abundance gradient. Our results are consistent with the O$^{++}$ electron-temperatures and absolute oxygen-abundances derived using optical direct-abundance methods by the CHemical Abundance Of Spirals (CHAOS) program, with a small difference ($\sim$ 1.5$\sigma$) in the radial abundance-gradients derived by the FIR/free-free-normalized vs. CHAOS/direct-abundance techniques. This initial result demonstrates the validity of the FIRA methodology --- with the promise of determining absolute metal-abundances within dusty star-forming galaxies, both locally and at high redshift.

\end{abstract}


\section{Introduction}

Tracing the buildup of metals through cosmic time is critical because of the monumental role metals play in the physics of the interstellar medium (ISM). Where present, even in minute quantities ($\sim$ 1-10:10,000 atoms per hydrogen atom), metals provide an efficient pathway for cooling the gas, and hence regulate the energy balance of the ISM. This metal-driven energy-regulation manifests itself in a myriad of ways, including in the initial-mass and luminosity of stellar populations \citep[e.g.,][]{Greif2015}. 

Increased metal content in the ISM also leads to increased dust content \citep[e.g.,][]{Remy-Ruyer2014}, which is not only interesting in its own right, but also affects measurements of the star-formation rate \citep[e.g., FIR-luminosity method;][]{Kennicutt1998}. Varying metal abundances also affect determinations of the molecular-gas mass in galaxies, which are generally made either using the Rayleigh-Jeans tail of the far-infrared (FIR) spectral-energy-distribution (SED) \citep[e.g.,][]{Scoville2016}, or through observations of the CO molecule, which is used as a proxy for the H$_2$ molecule. In particular, the CO-to-H$_2$ conversion-factor ($\alpha_{CO}$) varies as a function of metallicity \citep[e.g.,][]{Bolatto2013} because CO primarily relies on shielding from dust to survive ultra-violet (UV) irradiation, while H$_2$ molecules are largely self-shielding.

Furthermore, the metal abundance in galaxies can be used as a kind of ``cosmic clock'' to chart out their evolutionary histories. These metallicity clocks are sensitive to the buildup of successive generations of stars, as well as the infall of any pristine gas, from either galaxy mergers or from the cosmic web itself at high redshift, and outflows of metal-enriched gas launched by active galactic nuclei (AGN) and stellar feedback \citep[e.g.,][]{Maiolino2019}. 

Typically, gas-phase metal abundances in external galaxies are determined either using bright optical emission-lines, with a wide array of so-called ``strong-line'' calibrations that relate observed line-ratios to metal abundances \citep[e.g., R23, O32, N2O2;][]{Kobulnicky2004, Pilyugin2005, Diaz2000, KewleyDopita2002, Nagao2006, Perez-Montero2009}, or  alternatively, collisionally-excited ``direct-abundance" methods that directly determine the temperature and (sometimes) density of the line-emitting gas to calculate the line emissivities, which are used to relate the observed line-flux to an abundance \citep[e.g.,][]{Aller1954, Peimbert1967, Dinerstein1990, Pilyugin2006, Bresolin2009, Perez-Montero2014, Perez2015}. 

The advantages of the strong-line methods are that the observed spectral lines are bright, and hence easily detectable, and also in the optical part of the spectrum, such that a large number of extra-galactic sources have been observed in these lines \citep[e.g.,][]{Sanchez2012, Bundy2015, Blanton2017}. A major drawback of strong-line methods is the discrepancy in the derived abundances between the different calibrations, sometimes by as much as $\sim$\,0.8 dex \citep[e.g.,][]{KewleyEllison2008, Moustakas2010}, and the observed offset from the direct-abundances measurements. Also, with typical \HII-region electron-temperatures $\sim$\,8,000--15,000K, well below the energies of the strong-line emitting levels ($\sim$\,30,000\,K), the emissivities of these bright, collisionally-excited lines have an exponential sensitivity to the temperature of the emitting gas, making these strong-line abundance-diagnostics highly uncertain when temperature diagnostics are unavailable. 

Thus, the more fundamental abundance measurements are the direct-abundance methods that use collisionally-excited lines, where higher-lying, ``auroral", transitions ($E/k_B \sim$ 60,000\,K) are observed in conjunction with bright optical lines, to determine the electron temperature in the emitting gas. In that way, these temperature-corrected methods minimize the uncertainty in the line emissivity, and hence in the derived abundances. These methods still suffer somewhat from temperature uncertainties, in that the electron temperature must be measured for each ionization zone, or a temperature structure must be assumed for the emitting gas, and any fluctuations or sub-structure can affect the accuracy of the derived abundances \citep[e.g., $t^2$;][]{Peimbert1967}. Temperature and its (often unknown) structure therefore set the fundamental uncertainty of optical strong-line and direct-abundance measurements alike.

A promising alternative to optical-based metal-abundances that has the potential to side-step these fundamental, temperature-based, uncertainties is to employ collisionally-excited line transitions for which the excitation energy ($\lesssim$ 500\,K) is well \emph{below} the $k_B T$ energy of the electrons in the ionized gas, such that the line emissivities are largely insensitive to the temperature of the emitting regions \citep[e.g.,][]{Rubin1994}. This temperature agnosticism, similar to that of optical recombination-lines, which are typically too faint to be detected, is the primary motivation of the direct far-infrared abundance (FIRA) project introduced here. Instead of using \oIII\,5007\AA, for example, as a gas-phase oxygen-abundance indicator, we use the \oIII\,88\um\ line, which lies $E/k_B \,\lesssim\,200$\,K above ground. The \oIII\,88\um\ line also has the advantage that it is largely unaffected by dust extinction, which can be a great benefit, even in low-metallicity galaxies with significant star-formation rates \citep[$\gtrapprox$\,20\,M$_\odot$\,yr$^{-1}$; e.g.,][]{Reddy2015, Shivaei2020}. 

A drawback of using the \oIII\,88\um\ spectral line, and of far-infrared (FIR) fine-structure (FS) lines in general, as pointed out by \citet[][]{Maiolino2019}, is that its critical density \citep[510\,cm$^{-3}$,][]{Carilli2013} is well below that of most bright optical-lines. In the regime well below the critical density, the line emissivity, $j$, scales with density in exactly the same way as does the emissivity of the hydrogen-normalization required for the derivation of absolute abundances, $j \propto n_{e}^{2}$, such that the abundance determination is insensitive to electron density. Above the critical density, collisional de-excitation gains importance, so the line emissivity, $j$, transitions to scaling with $j \propto n_{e}$, deviating from the scaling of the hydrogen normalization and making the abundance determination density sensitive. We explore the severity of this density dependence on the derived abundances in this work, and compare the results obtained here using FIR direct-abundance techniques to those obtained with collisionally-excited optical-line direct-abundance and strong-line methods.

In addition to the \oIII\,88\um\ line, we need to employ some form of hydrogen normalization in order to obtain absolute abundances. Here we explore normalizations using both extinction-corrected H$\alpha$ recombination emission and radio free-free emission. Radio free-free emission, which has been used to determine gas-phase metallicities in galactic \HII\ regions \citep[e.g.,][]{Herter1981, Rudolph1997}, has the advantage of being independent of extinction. We compare these two hydrogen-normalizations, where both are observed, with the ultimate goal of calibrating radio free-free emission to determine absolute abundances in local ultra-luminous infrared galaxies (ULIRGs) and high-redshift dusty star-forming galaxies (DSFGs), where optical emission-lines are difficult or impossible to detect.

We also explore a mid-infrared based ionization correction-factor (ICF), employing the \neIII\,15.6\um\ and \neII\,12.8\um\ lines, which accounts for the ionization states of oxygen that are unobservable in the far-infrared. Such mid-IR-derived ICFs have been developed previously in the literature for, e.g., neon and sulfur \citep[e.g.,][]{Dors2013, Dors2016, Armah2021}.

The power of the FIRA project lies in bringing together all of the datasets necessary to put direct FIR-based absolute-abundances on a firm footing with similar studies in the optical. Specifically, we combine observations of the \oIII\,88\um\ line from the Herschel/Photodetector Array Camera and Spectrometer (PACS), the \neIII\,15.6\um\ and \neII\,12.8\um\ lines from the Spitzer/Infrared Spectrograph (IRS), and the filled integral-field unit (IFU) H$\alpha$, H$\beta$, \oIII\,5007\AA, \sII\,6716\AA\ and 6730\AA\ maps from the Potsdam Multi-Aperture Spectrograph (PMAS) in PMAS fiber Package (PPAK) mode, to determine the electron temperatures and densities within the emitting gas, as well as to determine the O$^{++}$ ionic abundances, ICFs, and total oxygen-abundances within the targeted ionized-gas regions. 

As the initial target for this FIRA survey, we explore M101, which has been the subject of many chemical-abundance studies \citep[e.g.,][]{Rayo1982, Kennicutt2003, Li2013}. Critically, M101 is a prominent target of both the Star-Formation in Radio Survey \citep[SFRS,][]{Linden2020, Murphy2018}, which mapped \HII\ regions in multi-band radio-continuum with the Karl G. Jansky Very Large Array (VLA), and the CHemical Abundances Of Spirals \citep[CHAOS,][]{Berg2015, Croxall2016, Berg2020} project, with direct, auroral-line abundance-determinations in $\sim$100 \HII\ regions across the galaxy, and excellent overlapping observations with the FIR, radio, and optical observations necessary for the FIR/radio and FIR/optical abundance-methods explored here. The CHAOS observations allow us to compare our FIR/radio and FIR/optical abundance-determinations to the gold-standard in direct collisionally-excited optically-derived abundances.

This paper is organized as follows. In Section 2, we describe the target selection, observations, and data reduction that comprise this paper. In Section 3 we derive the physical parameters, electron temperature and density, in the targeted \HII\ regions of M101, comparing them to those calculated by the CHAOS group, using collisionally-excited optical auroral-line methods. In Section 4, we explore the choice of hydrogen normalization, H$\alpha$ or radio free-free emission, and its effect on the derived abundances. In Section 5 we derive the O$^{++}$ ionic abundances in the targeted \HII\ regions of M101, comparing them to those calculated by CHAOS. In Section 6 we derive a mid-infrared ionization correction factor, using photoionization models, to account for the O$^+$ ion, which cannot be observed in the far-infrared. In Section 7 we present the absolute O/H abundances in the targeted \HII\ regions of M101, comparing them to those derived by the CHAOS collaboration, as well as strong-line methods. In Sections 8 and 9 we discuss and summarize, respectively, our main findings from this first paper of the FIRA project.

\begin{figure*}[t]
\begin{center}		
\includegraphics[width={0.95\linewidth}]{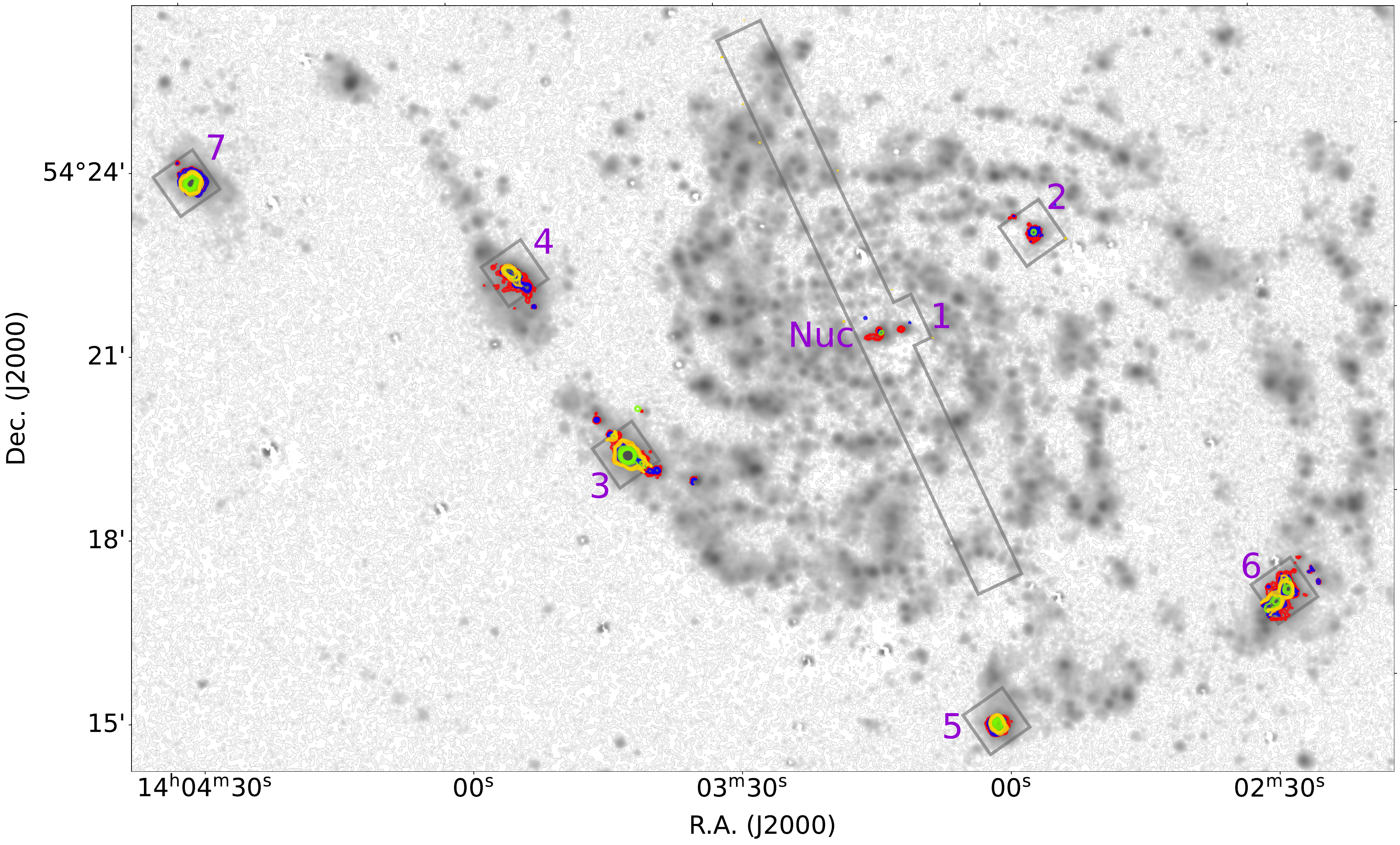}
\caption{Overview of M101 observations. Herschel/PACS \oIII\,88\um\ (yellow), VLA 33GHz radio continuum (green) from SFRS \citep{Linden2020}, and PPAK IFU H$\alpha$ (red) and \oIII\,5007\AA\ (blue), contours are overplotted on a greyscale background-subtracted H$\alpha$ narrow-band image from \cite{Hoopes2001}. The faint grey outlines denote the footprint of the PACS \oIII\,88\um\ observations. The numbers indicate the naming scheme used to delineate the targeted regions in this paper (see Table \ref{Region_Coordinates}).}
\label{M101_Overview_Figure}
\end{center}
\end{figure*}

\section{Observations and Data Reduction}

\subsection{Target Selection}

M101, a local \citep[D = 7.4 Mpc,][]{Ferrarese2000Distances}, nearly face-on spiral, is a very suitable first target for the FIRA project, as it has excellent overlapping coverage in the spectral lines and radio continuum necessary for this project, including the \oIII\,88\um\ line from Herschel/PACS, the \neIII\,15.6\um\ and \neII\,12.8\um\ lines from Spitzer/IRS, the H$\alpha$, H$\beta$, \oIII\,5007\AA, \sII\,6716\AA\ and 6730\AA\ lines from PPAK, and 33GHz VLA maps.

The targeted \HII\ regions were selected based on bright \oIII\,88\um\ line-detections from Herschel/PACS, as well as overlapping observations with the VLA, PPAK, and Spitzer. The targeted coordinates were adopted from the SFRS survey and the naming nomenclature follows that described in Section 3.2 of \cite{Linden2020}. In addition to the nucleus of M101, we target 7 extra-nuclear \HII\ regions, out to a de-projected galacto-centric radius of $\sim$ 24 kpc (see Table \ref{Region_Coordinates} and Figure \ref{M101_Overview_Figure}).

\subsection{Herschel/PACS}

The \oIII\,88\um\ line was observed in M101 using PACS \citep[][]{Poglitsch2010} on board the Herschel Space Observatory\footnote{\emph{Herschel} is an ESA space observatory with science instruments provided by European-led Principal Investigator consortia and with important participation from NASA.} \citep{Pilbratt2010}, as part of the Key Insights on Nearby Galaxies: a Far-Infrared Survey with Herschel (KINGFISH) survey \citep{Kennicutt2011}, on 2011 June 26, 2011 July 01, and 2011 July 05 (OBSIDs: 1342223146, 1342223390, 1342223392, 1342223394, 1342223396, 1342223718, 1342223720, 1342223721).

The PACS IFU is comprised of a 5$\times$5 array of \emph{spaxels} (spatial-pixels), each of which covers 9.4\arcsec\ $\times$ 9.4\arcsec\ on the sky, for a total field-of-view of $\sim$ 47\arcsec\ $\times$ 47\arcsec. The observations of M101 were conducted using the un-chopped mapping mode, wherein the detector footprint is raster scanned across the target galaxy, increasing the sampling of the point-spread function (PSF), which at 88\um\ ($\sim$ 9\arcsec) is under-sampled by the 9.4\arcsec\ pixels.

These observations were reduced using the Herschel Interactive Processing Environment (HIPE), version 15.0.1 \citep{Ott2010}, following the methods of the KINGFISH DR3 release\footnote{https://irsa.ipac.caltech.edu/data/Herschel/KINGFISH/docs\\/KINGFISH\_DR3.pdf}. The output of this reduction is very similar to that available in the KINGFISH data repository: spectral cubes of the \oIII\,88\um\ line with $\sim$ 2.1\arcsec\ $\times$ 2.1\arcsec\ spatial pixels and $\sim$ 60 km s$^{-1}$ velocity bins, which have been up-sampled from the spectrometer's native resolution of $\sim$ 120 km s$^{-1}$ at 88 $\micron$. The re-reduction of these data was necessary to include an updated understanding of the PACS calibration for extended sources, the net result of which boosts the flux at 88 $\micron$ relative to the KINGFISH data release by $\sim$ 30\%\footnote{http://herschel.esac.esa.int/twiki/bin/view/Public\\/HipeWhatsNew14x\#Spectroscopy\_AN1}. Finally, we convolve the PACS maps with a Gaussian convolution kernel, to obtain a final spatial resolution of $\sim$ 10\arcsec. Figure \ref{M101_Regions_Figure_First} presents the Herschel/PACS, VLA 33 GHz, PPAK, Spitzer, and CHAOS observations used in this analysis.

\subsection{VLA}

Radio-continuum observations of M101 were carried out with the VLA\footnote{The National Radio Astronomy Observatory is a facility of the National Science Foundation operated under cooperative agreement by Associated Universities, Inc.} as part of the SFRS survey, at 3, 15, and 33\,GHz \citep{Linden2020}. This multi-band data-set allows the radio spectral-energy-distribution (SED) to be decomposed into free-free and synchrotron components, critical to our goal of using the free-free emission as a hydrogen normalization for the gas-phase abundance determination.

The measurement-sets of the 33\,GHz VLA observations were each imaged at $\sim$ 2\arcsec\ resolution, then convolved with a Gaussian kernel to a $\sim$ 10\arcsec\ beam, for comparison at common resolution with the other data-sets used in this paper.

\subsection{PPAK}

M101 was observed using the PMAS spectrograph \citep{Roth2005} in PPAK mode \citep{Verheijen2004, Kelz2006} on the 3.5m Calar Alto telescope on 2014 April 13-14. This instrument covers a hexagonal $\sim$1\arcmin\ diameter field-of-view with 331 science fibers, 2.7\arcsec\ in diameter, that achieve a 60\% filling factor. We used the V300 grating to observe from 3650-7000\AA\ at $\sim$ 180\,km\,s$^{-1}$ spectral resolution, sufficient to deblend the [\ion{S}{2}] doublet as well as deblend H$\alpha$ from the neighboring [\ion{N}{2}] doublet.  We observed in eight positions within M101, selected to cover the galaxy center as well as seven extra-nuclear regions that had been previously targeted with Herschel/PACS as part of the KINGFISH project. To achieve 100\% filling factor, we observe each pointing in three dither positions, for a total on-source integration time of 1 hour. As the target galaxy fills the entire field of view, dedicated sky observations were taken before and after each science exposure with 20m integration times.

Our data reduction is carried out following the steps outlined in \citet{Kreckel2013}, and summarized here. All data reduction steps are executed using the p3d package \citep{Sandin2010}, v2.2.5.1 `Serenity'. Arc and calibration lamp images are taken at each science position, and used to derive a wavelength solution and trace the dispersed spectra. Spectrophotometric standard stars were observed at the beginning and end of each night, and used to calibrate each science exposure. The resulting calibrated spectra are fit using a combination of stellar population templates \citep{BC03} and Gaussian profile emission line fits, along with an 8th order multiplicative polynomial to account for calibration inaccuracies. As most of these regions are at large galactocentric radii, the stellar continuum is generally not significantly detected against the sky background.  All spectral fitting is carried out using the software packages ppxf \citep{Cappellari2004} and gandalf \citep{Sarzi2006}. This work makes use of the resulting line maps for the strong lines H$\beta$, [\ion{O}{3}]$\lambda$5007, H$\alpha$, [\ion{N}{2}]$\lambda$6583, [\ion{S}{2}]$\lambda$6717 and [\ion{S}{2}]$\lambda$6731.

An astrometric solution is determined by comparing to ground-based H$\alpha$ narrow-band image available on the NASA/IPAC Extragalactic Database (NED) from \citet{Hoopes2001}. A slight offset in the archival image is apparent in relation to SDSS g-band imaging \citep{Alam2015}, so we apply an intial correction by aligning (by eye) the positions of stars between the two images. This is accurate to $<$\,1\arcsec, well below the size of the apertures extraced for this paper. We align our PPAK H$\alpha$ images with this corrected archival H$\alpha$ narrow-band image using a 2D cross correlation, via the idl task ``CORREL\_IMAGES". 

For all maps, we correct for foreground Milky Way extinction, assuming E(B-V)=0.008 mag \citep{Schlafly2011}, and internal reddening, assuming the Balmer decrement and R$_V$=3.1, and applying a \cite{ccm} extinction law. The data were then convolved with a Gaussian kernel to a $\sim$ 10\arcsec\ beam to match the other data-sets used in this paper.

\subsection{Spitzer/IRS}

M101 was observed using the InfraRed Spectrograph \citep[IRS,][]{Houck2004} on board the Spitzer Space Telescope. The \neII\,12.8\um\ and \neIII\, 15.6\um\ lines were observed with the short-wavelength, high-resolution (R $\sim$ 600) module, SH (AORKEYs= 4372224, 4372480, 4372736, 4372992; P.I.: George Rieke). The data were reduced with CUBISM \citep{Smith2007} and spectra for the two neon lines were extracted from the entire footprint of the SH slit in each pointing (except for region 6a, see Figure \ref{M101_Regions_Figure_First}). For region 6a, where a portion of the Spitzer/IRS high-res field-of-view is located outside of the 20\arcsec\ extraction-aperture, we consider only the portion of the flux contained within the 20\arcsec\ extraction aperture. The resulting spectra were numerically integrated to obtain the flux in each line.

The [Ne\,{\sc iii}] 15.6 $\micron$, [Ne\,{\sc ii}] 12.8 $\micron$, and [S\,{\sc iii}] 18.7 and 33.5 $\micron$ lines were also observed with the low-resolution (R $\sim$ 100) module \citep{Gordon2008}; however, the blending of the \neII\ line with a nearby PAH emission-feature makes these line-flux measurements quite uncertain. Hence, we use the high-resolution data for the analysis presented here. This blending is less of a concern with the \sIII\ lines.

Where overlapping observations exist with the other data-sets presented here, we use self-contained line-ratios from Spitzer/IRS to constrain the electron density and ionization-correction factor within the \HII\ regions of M101. We do not combine the Spitzer/IRS flux values with those measured using other observatories.

\subsection{LBT/MODS CHAOS}

Highly-sensitive optical-slit-spectroscopy of M101 was obtained as part of the CHAOS program \citep{Berg2015, Croxall2016}, a project with the goal of determining gas-phase absolute-metallicities using collisionally-excited line direct-abundance methods in a large sample of nearby spiral-galaxies. These observations, with the ability to measure the \oIII\,4363\AA\ auroral line and \oIII\,5007\AA\ line in dozens of extra-galactic \HII\ regions simultaneously, probe the electron temperature in the targeted \HII\ regions.

For the M101 observations, extraction slits over the range 1.2$\arcsec$$\times$4$\arcsec$ to 1.2$\arcsec$$\times$13$\arcsec$, corresponding to $\sim$ 31$\times$104\,pc and 31$\times$338\,pc, respectively, were employed. We compare the \HII\ region physical conditions, ionic abundances, and absolute abundances derived using the far-IR direct-abundance method against those derived by CHAOS, wherever overlapping observations exist. We do not combine the CHAOS flux values with those measured using the other observatories.

\begin{deluxetable}{cccc}
\tablecaption{Targeted Regions Within M101}
\tablecolumns{4}
\tablenum{1}
\tablehead{
\colhead{Region} & \colhead{R.A.} & \colhead{Dec.} & \colhead{R$_G$\tablenotemark{a}}\\
\colhead{} & \colhead{} & \colhead{} & \colhead{(kpc)}\\
}
\startdata
Nuc & 14h03m12.531s & +54d20m55.200s & 0.062 \\
1 & 14h03m10.200s & +54d20m57.800s & 0.744 \\
2 & 14h02m55.000s & +54d22m27.500s & 6.451 \\
3a & 14h03m38.317s & +54d18m51.398s & 9.307 \\
3b & 14h03m39.894s & +54d18m56.799s & 9.655 \\
3c & 14h03m41.437s & +54d19m04.900s & 9.964 \\
3d & 14h03m42.912s & +54d19m24.669s & 10.113 \\
4a & 14h03m52.036s & +54d21m52.500s & 12.149 \\
4b & 14h03m52.997s & +54d21m57.300s & 12.453 \\
4c & 14h03m53.203s & +54d22m06.300s & 12.540 \\
4d & 14h03m53.993s & +54d22m10.800s & 12.795 \\
5 & 14h03m01.203s & +54d14m28.400s & 13.129 \\
6a & 14h02m28.203s & +54d16m27.200s & 15.707 \\
6b & 14h02m29.607s & +54d16m15.799s & 15.550 \\
6c & 14h02m30.566s & +54d16m09.798s & 15.422 \\
7 & 14h04m29.334s & +54d23m47.600s & 23.861 \\
\enddata
\tablecomments{The targeted \HII\ regions within M101 (shown in Figures \ref{M101_Overview_Figure} and \ref{M101_Regions_Figure_First}). R$_G$ denotes the de-projected galacto-centric radius from the center of M101.}
\label{Region_Coordinates}
\tablenotetext{a}{\cite{Linden2020}}
\end{deluxetable}

\subsection{Flux Extraction}
After the PACS, PPAK, and VLA maps were all convolved with a Gaussian kernel to a common resolution of $\sim$ 10\arcsec, we performed a relative astrometric correction between the images, translating the different maps into the VLA reference frame, before extracting the flux from 20\arcsec-diameter apertures, centered on the coordinates listed in Table \ref{Region_Coordinates} (and see Figure \ref{M101_Regions_Figure_First}). This larger aperture, 2$\times$ the convolved common-resolution, is chosen to avoid any slight deviations from the nominal common-resolution of 10\arcsec\ in any of the maps. 

In addition to the statistical error that is propagated through the line-flux and continuum-flux calculations, we add, in quadrature, an estimated absolute flux-calibration uncertainty of 10\% to the PACS, PPAK, and VLA observations, necessary for taking the ratios of lines observed with different instruments. The resulting line-flux and continuum measurements, as well as their associated uncertainties, can be found in Table \ref{Flux_Table}.

We do not make any attempt to combine the CHAOS/MODS flux values, extracted from $\sim$ 1\arcsec-wide slits, nor the Spitzer/IRS measurements, with fluxes extracted from the larger 20\arcsec\ apertures of the other maps. We believe this is appropriate because Herschel/PACS, PPAK, and the VLA produce filled maps, which we can convolve to a common resolution before extracting flux from a common aperture, whereas this is not possible with the slits of CHAOS/MODS and Spitzer/IRS. We simply compare the physical conditions of the \HII\ regions derived in the small CHAOS slits (e.g., electron temperature, O$^{++}$ ionic-abundance, and absolute O/H-abundance), and the moderate Spitzer slits (e.g., electron density and ionization parameter), to those measured in the larger overlapping 20\arcsec\ regions.

\begin{deluxetable*}{ccccccccc}
\tabletypesize{\scriptsize}
\tablecaption{Spectral line and continuum observations of M101}
\tablecolumns{9}
\tablenum{2}
\tablewidth{0pt}
\tablehead{
\colhead{Region} & \colhead{\oIII\,88\um} & \colhead{\oIII\,5007\AA} & \colhead{H$\alpha$} & \colhead{\sII\,6716\AA} & \colhead{\sII\,6730\AA} & \colhead{I$_{\rm 33GHz}$} & \colhead{f$_{\rm free-free}$\tablenotemark{a}} & \colhead{I$_{\rm free-free}$}\\
\colhead{} & \colhead{(10$^{-13}$\,erg\,s$^{-1}$\,cm$^{-2}$)} & \colhead{(10$^{-13}$\,erg\,s$^{-1}$\,cm$^{-2}$)} & \colhead{(10$^{-13}$\,erg\,s$^{-1}$\,cm$^{-2}$)} & \colhead{(10$^{-13}$\,erg\,s$^{-1}$\,cm$^{-2}$)} & \colhead{(10$^{-13}$\,erg\,s$^{-1}$\,cm$^{-2}$)} & \colhead{(mJy)} & \colhead{-} & \colhead{(mJy)}\\
}
\startdata
Nuc & 0.49 \,$\pm$\, 0.05 & 0.09 \,$\pm$\, 0.01 & 3.03 \,$\pm$\, 0.30 & 0.26 \,$\pm$\, 0.03 & 0.18 \,$\pm$\, 0.02 & 0.42 \,$\pm$\, 0.08 & 0.77 \,$\pm$\, 0.03 & 0.32 \,$\pm$\, 0.06 \\
1 & 0.33 \,$\pm$\, 0.04 & 0.01 \,$\pm$\, 0.01 & 1.23 \,$\pm$\, 0.12 & 0.06 \,$\pm$\, 0.01 & 0.05 \,$\pm$\, 0.01 & 0.13 \,$\pm$\, 0.02 & 0.97 \,$\pm$\, 0.04 & 0.13 \,$\pm$\, 0.02 \\
2 & 0.74 \,$\pm$\, 0.08 & 0.68 \,$\pm$\, 0.07 & 9.48 \,$\pm$\, 0.95 & 1.33 \,$\pm$\, 0.13 & 0.96 \,$\pm$\, 0.10 & 0.38 \,$\pm$\, 0.05 & 1.00 \,$\pm$\, 0.02 & 0.38 \,$\pm$\, 0.05 \\
3a & ... & 3.52 \,$\pm$\, 0.35 & 7.85 \,$\pm$\, 0.78 & 1.07 \,$\pm$\, 0.11 & 0.76 \,$\pm$\, 0.08 & 0.25 \,$\pm$\, 0.21 & 0.82 \,$\pm$\, 0.09 & 0.20 \,$\pm$\, 0.18 \\
3b & ... & 10.15 \,$\pm$\, 1.02 & 17.12 \,$\pm$\, 1.71 & 2.13 \,$\pm$\, 0.21 & 1.55 \,$\pm$\, 0.16 & 1.44 \,$\pm$\, 0.26 & 1.00 \,$\pm$\, 0.02 & 1.44 \,$\pm$\, 0.26 \\
3c & 40.39 \,$\pm$\, 4.04 & 60.26 \,$\pm$\, 6.03 & 61.27 \,$\pm$\, 6.13 & 5.01 \,$\pm$\, 0.50 & 3.88 \,$\pm$\, 0.39 & 6.65 \,$\pm$\, 0.70 & 0.97 \,$\pm$\, 0.01 & 6.45 \,$\pm$\, 0.68 \\
3d & ... & 3.84 \,$\pm$\, 0.38 & 6.62 \,$\pm$\, 0.66 & 0.87 \,$\pm$\, 0.09 & 0.64 \,$\pm$\, 0.06 & 0.47 \,$\pm$\, 0.22 & 0.86 \,$\pm$\, 0.04 & 0.40 \,$\pm$\, 0.19 \\
4a & ... & 5.82 \,$\pm$\, 0.58 & 6.49 \,$\pm$\, 0.65 & 0.71 \,$\pm$\, 0.07 & 0.44 \,$\pm$\, 0.04 & 0.35 \,$\pm$\, 0.10 & 0.91 \,$\pm$\, 0.03 & 0.32 \,$\pm$\, 0.09 \\
4b & 4.56 \,$\pm$\, 0.46 & 9.52 \,$\pm$\, 0.95 & 10.55 \,$\pm$\, 1.06 & 1.18 \,$\pm$\, 0.12 & 0.81 \,$\pm$\, 0.08 & 0.71 \,$\pm$\, 0.12 & 0.97 \,$\pm$\, 0.02 & 0.69 \,$\pm$\, 0.12 \\
4c & 5.98 \,$\pm$\, 0.60 & 12.45 \,$\pm$\, 1.24 & 13.83 \,$\pm$\, 1.38 & 1.49 \,$\pm$\, 0.15 & 1.07 \,$\pm$\, 0.11 & 0.96 \,$\pm$\, 0.14 & 0.93 \,$\pm$\, 0.01 & 0.89 \,$\pm$\, 0.13 \\
4d & 5.41 \,$\pm$\, 0.54 & 10.57 \,$\pm$\, 1.06 & 12.03 \,$\pm$\, 1.20 & 1.25 \,$\pm$\, 0.13 & 0.88 \,$\pm$\, 0.09 & 0.81 \,$\pm$\, 0.13 & 0.86 \,$\pm$\, 0.02 & 0.70 \,$\pm$\, 0.11 \\
5 & 9.76 \,$\pm$\, 0.98 & 26.33 \,$\pm$\, 2.63 & 23.75 \,$\pm$\, 2.37 & 2.39 \,$\pm$\, 0.24 & 1.82 \,$\pm$\, 0.18 & 1.81 \,$\pm$\, 0.19 & 0.98 \,$\pm$\, 0.01 & 1.78 \,$\pm$\, 0.19 \\
6a & 8.89 \,$\pm$\, 0.89 & 18.85 \,$\pm$\, 1.89 & 18.54 \,$\pm$\, 1.85 & 1.77 \,$\pm$\, 0.18 & 1.31 \,$\pm$\, 0.13 & 1.12 \,$\pm$\, 0.12 & 0.91 \,$\pm$\, 0.01 & 1.02 \,$\pm$\, 0.11 \\
6b & 9.50 \,$\pm$\, 0.95 & 17.10 \,$\pm$\, 1.71 & 19.19 \,$\pm$\, 1.92 & 2.44 \,$\pm$\, 0.24 & 1.76 \,$\pm$\, 0.18 & 1.32 \,$\pm$\, 0.14 & 0.83 \,$\pm$\, 0.02 & 1.10 \,$\pm$\, 0.12 \\
6c & ... & 13.10 \,$\pm$\, 1.31 & 14.23 \,$\pm$\, 1.42 & 1.82 \,$\pm$\, 0.18 & 1.32 \,$\pm$\, 0.13 & 1.07 \,$\pm$\, 0.11 & 0.91 \,$\pm$\, 0.02 & 0.97 \,$\pm$\, 0.10 \\
7 & 13.62 \,$\pm$\, 1.36 & 81.11 \,$\pm$\, 8.11 & 39.62 \,$\pm$\, 3.96 & 2.11 \,$\pm$\, 0.21 & 1.63 \,$\pm$\, 0.16 & 3.54 \,$\pm$\, 0.37 & 0.88 \,$\pm$\, 0.01 & 3.12 \,$\pm$\, 0.32 \\
\enddata
\tablecomments{The observed PACS \oIII\,88\um, (extinction-corrected) PPAK \oIII\,5007\AA, H$\alpha$, \sII\,6716\AA, and \sII\,6730\AA\ line-fluxes, as well as the VLA 33\,GHz continuum-flux measurements in the targeted regions of M101. Extraction apertures are 20\arcsec\ in diameter, centered on the coordinates listed in Table \ref{Region_Coordinates}. For the extraction apertures not completely contained within the PACS footprint, no \oIII\,88\um\ line-flux is reported. f$_{free-free}$ denotes the free-free fraction of the observed continuum flux at 33\,GHz, such that the free-free flux is given by: I$_{\rm free-free}$ $=$ f$_{\rm free-free}$ $\times$ I$_{\rm 33\,GHz}$.}
\tablenotetext{a}{\cite{Linden2020}}
\label{Flux_Table}
\end{deluxetable*}

\begin{figure*}
\begin{center}
\begin{tabular}{c}
\includegraphics[width=0.8\textwidth]{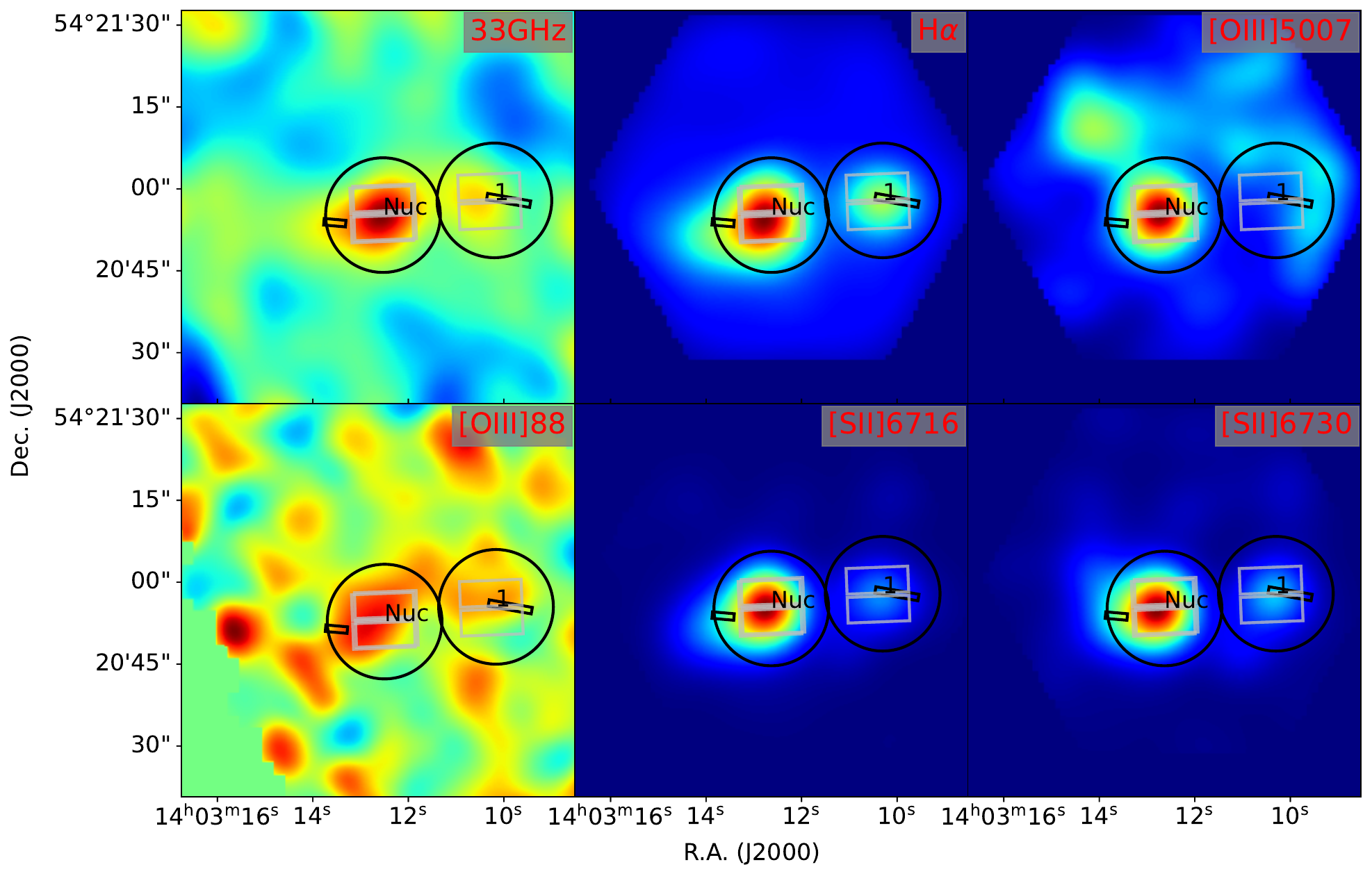} \\
\includegraphics[width=0.8\textwidth]{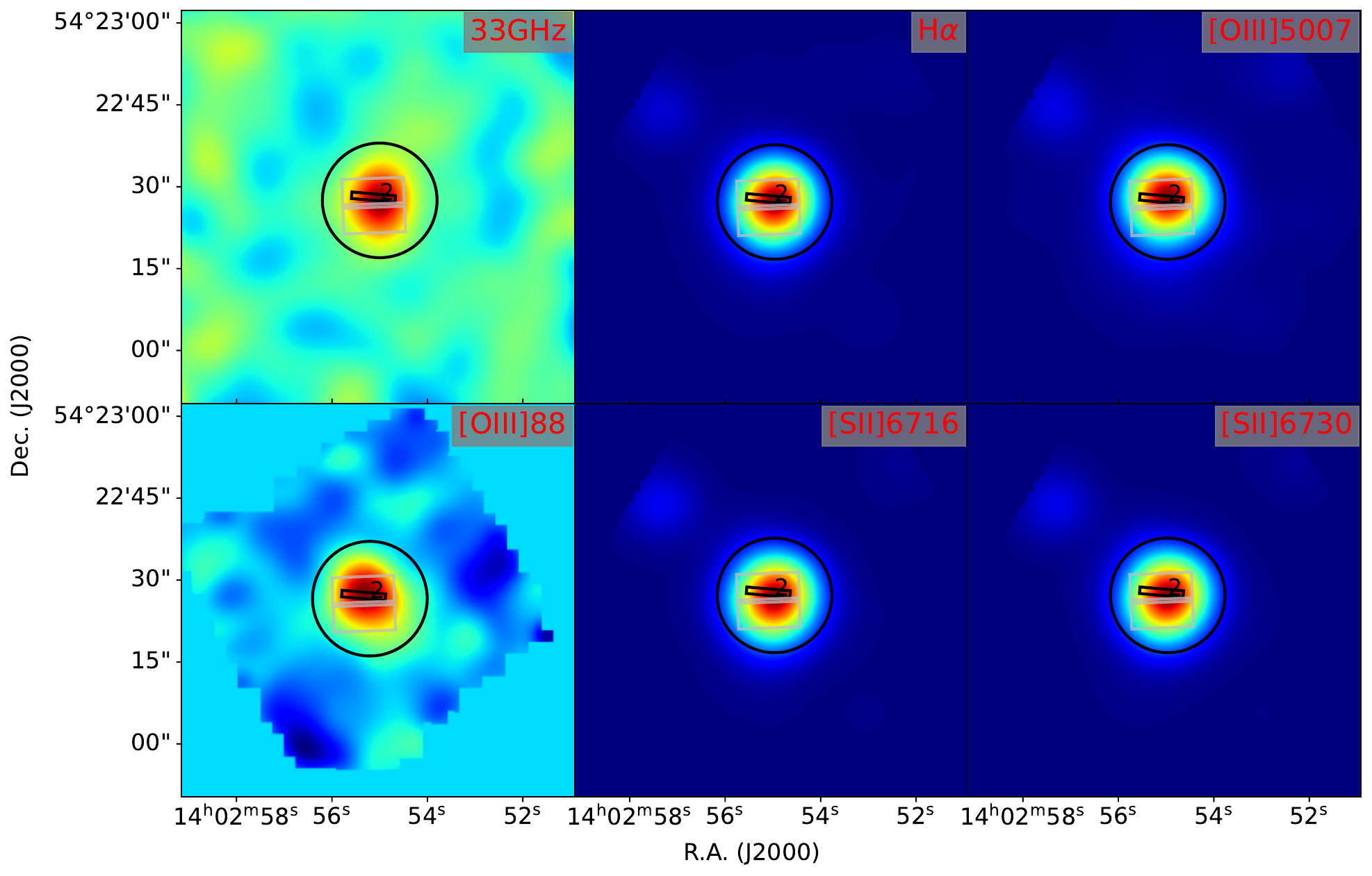}
\end{tabular}
\end{center}
\end{figure*}

\begin{figure*}
\begin{center}
\begin{tabular}{c}
\includegraphics[width=0.8\textwidth]{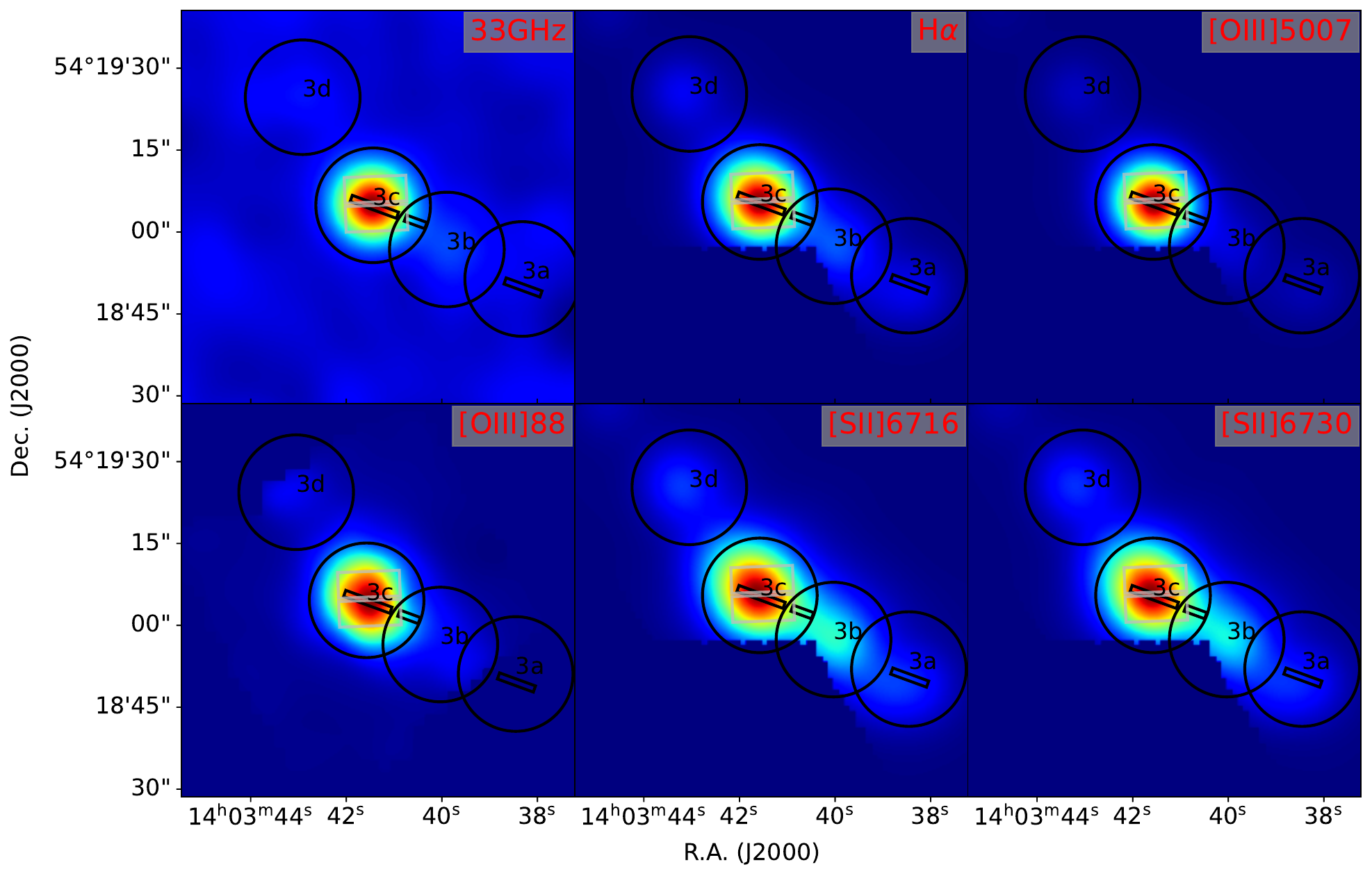} \\
\includegraphics[width=0.8\textwidth]{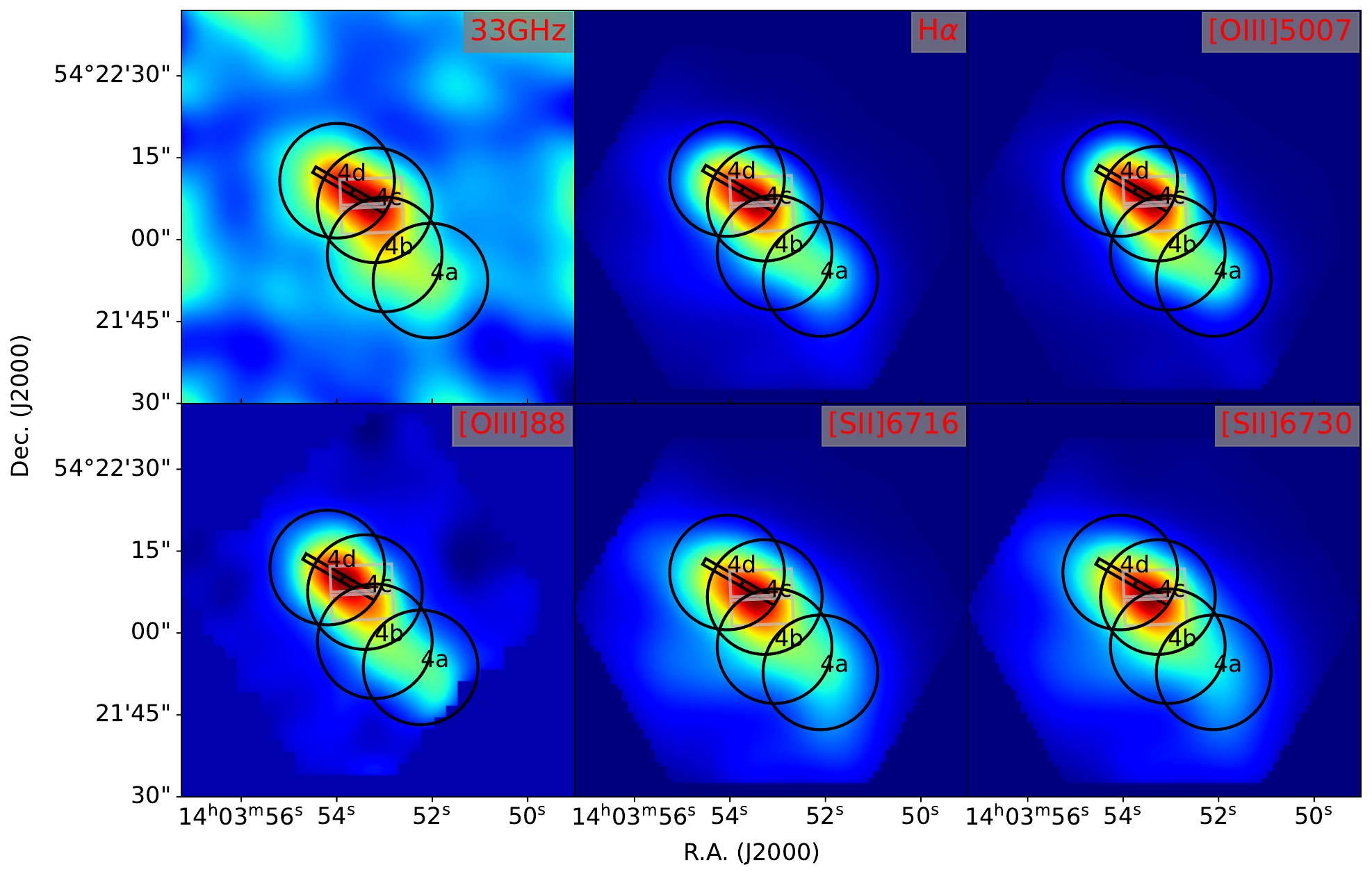}
\end{tabular}
\end{center}
\end{figure*}

\begin{figure*}
\begin{center}
\begin{tabular}{c}
\includegraphics[width=0.8\textwidth]{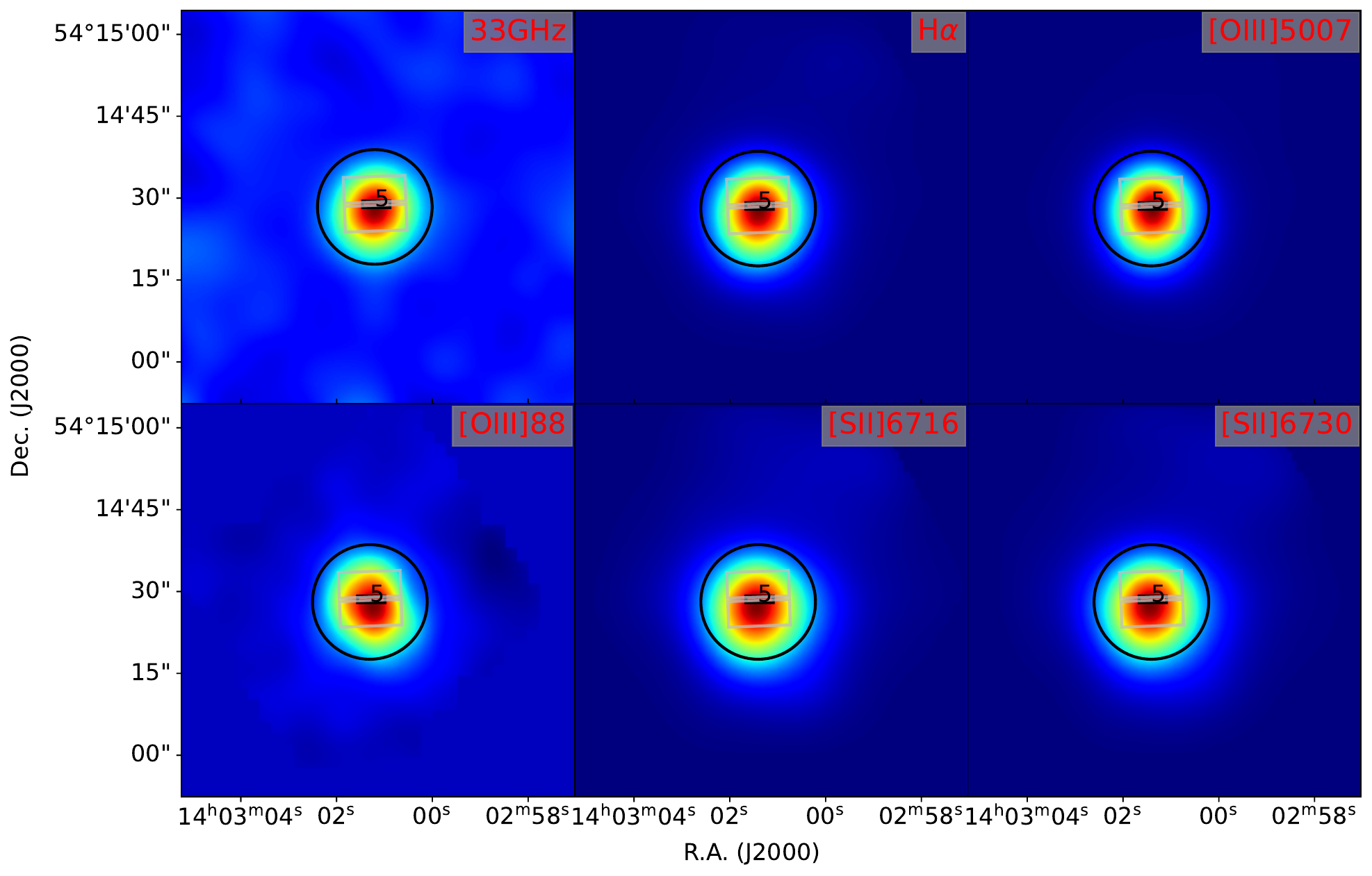} \\
\includegraphics[width=0.8\textwidth]{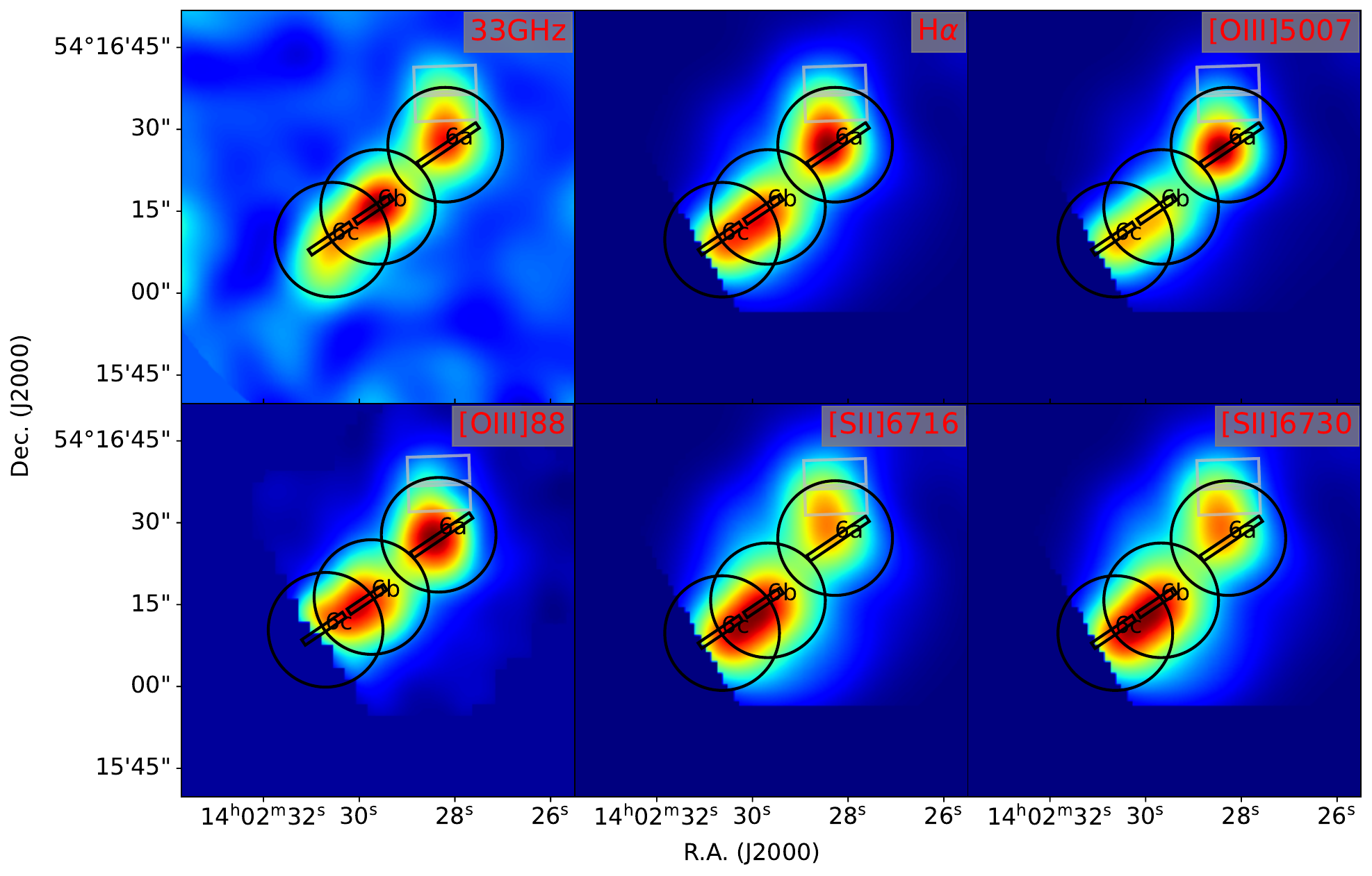}
\end{tabular}
\end{center}
\end{figure*}

\begin{figure*}
\begin{center}
\begin{tabular}{c}
\includegraphics[width=0.8\textwidth]{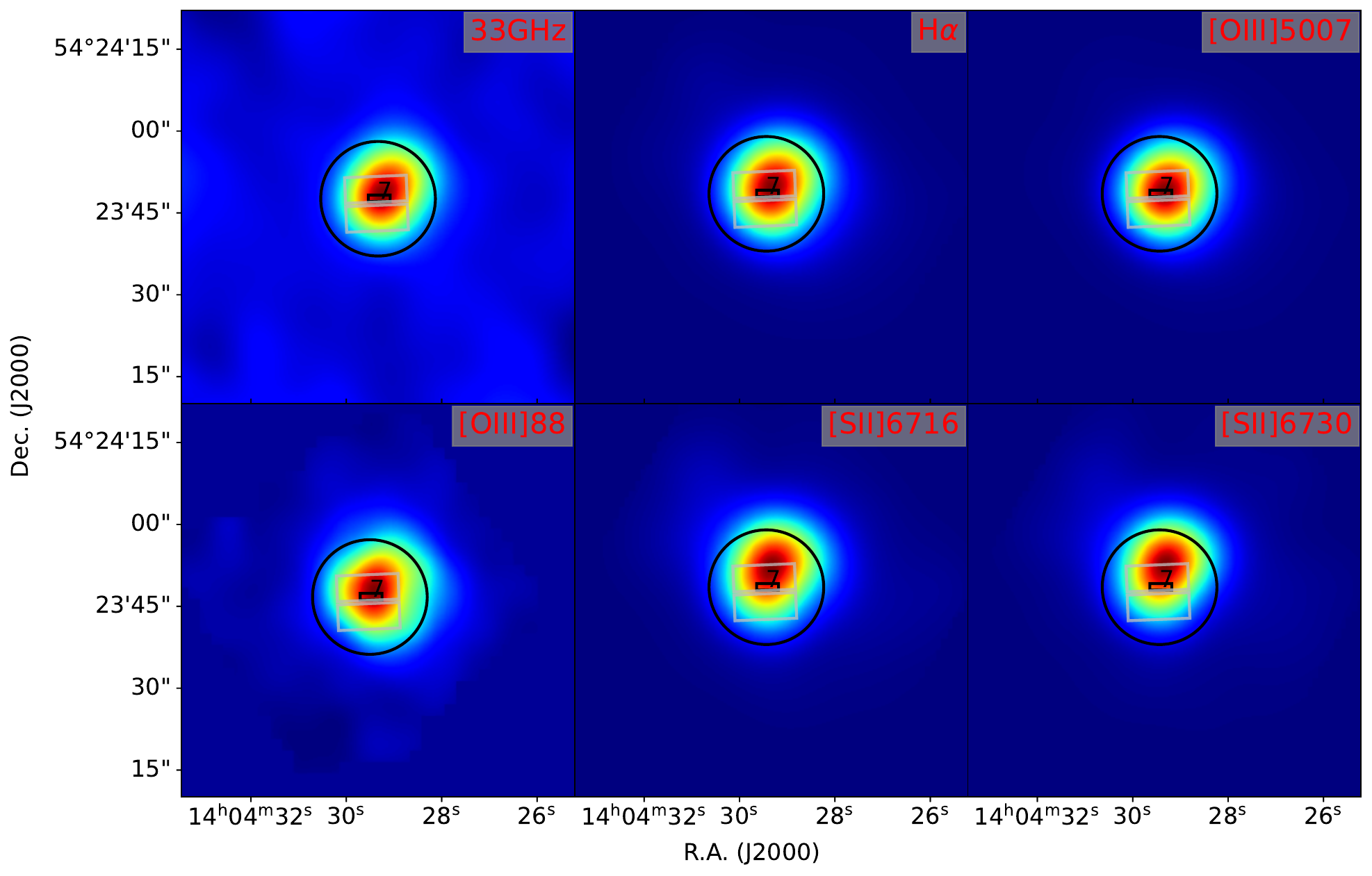}
\end{tabular}
\caption{(Continued from previous pages) Moment-zero maps of the spectral lines (presented here), and radio continuum \citep[from the SFRS survey;][]{Linden2020}, observed in the targeted regions of M101. Black circles denote the 20\arcsec\ extraction-regions employed here, while the smaller black rectangles and intermediate grey rectangles denote the CHAOS/MODS and Spitzer/IRS extraction-slits, respectively, where they overlap with our targeted regions.}
\label{M101_Regions_Figure_First}
\end{center}
\end{figure*}

\section{Density \& Temperature Determinations}

Far-IR FS lines are excellent probes of the ISM, their emissivity being largely temperature insensitive and their propagation largely unaffected by dust attenuation. While the emissivity of the far-IR FS lines is not strongly affected by the electron temperature within the emitting regions, it is affected by the electron density within those regions. In the low-density limit, well below the critical density of the line, the line-emissivity per-unit-volume (or line-emissivity per-collisional-pair), $\epsilon \equiv j/(n_e n_{\mathrm{ion}})$, is approximately constant with density, or equivalently $j$ $\propto$ $n_{e}^{2}$, such that varying density among or within \HII\ regions does not affect the abundances derived in these regions. While in the high-density limit, the line-emissivity per-unit-volume exhibits an inverse-linear dependence on the electron density, $\epsilon \propto n_{e}^{-1}$, or equivalently $j \propto n_{e}$. In practice, much of the ionized gas in the Milky Way and in extra-galactic sources is found to be at low density, \citep[e.g.,][]{Goldsmith2015, Herrera-Camus2016Density}; however, we do not make any assumptions about the gas-density in the following analysis, as our observations allow us to determine the density directly.

In addition to being unaffected by dust attenuation and variations in electron temperature, when combined with optical nebular lines, the FIR-FS lines provide a method of directly measuring the temperatures in ionized-gas regions. This eliminates the need for challenging observations of inherently faint auroral-lines because optical strong-lines, \oIII\,5007\AA\ for example, can effectively serve as the `auroral' lines for the FIR-FS lines, probing the temperature distribution of electrons closer to its peak (typically $\sim$ 10,000\,K). The introduction of the \oIII\,5007\AA\ line does make this FIR/optical temperature-diagnostic moderately sensitive to dust attenuation; however, it is not nearly as sensitive as is the equivalent \oIII\,4363\AA/5007\AA\ temperature diagnostic.

\begin{figure}
\begin{center}		
\includegraphics[width=0.45\textwidth]{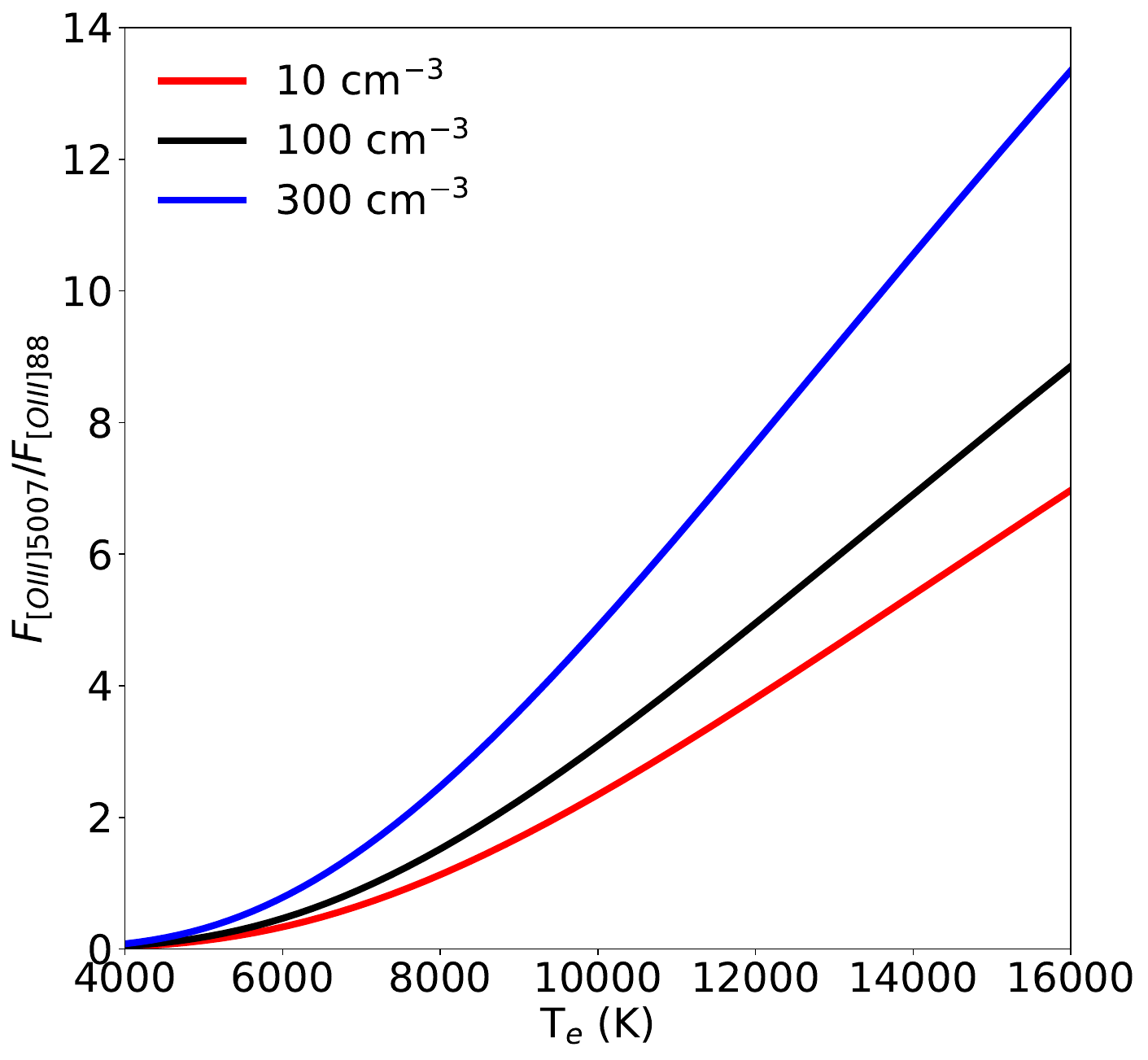}{(a)}
\includegraphics[width=0.45\textwidth]{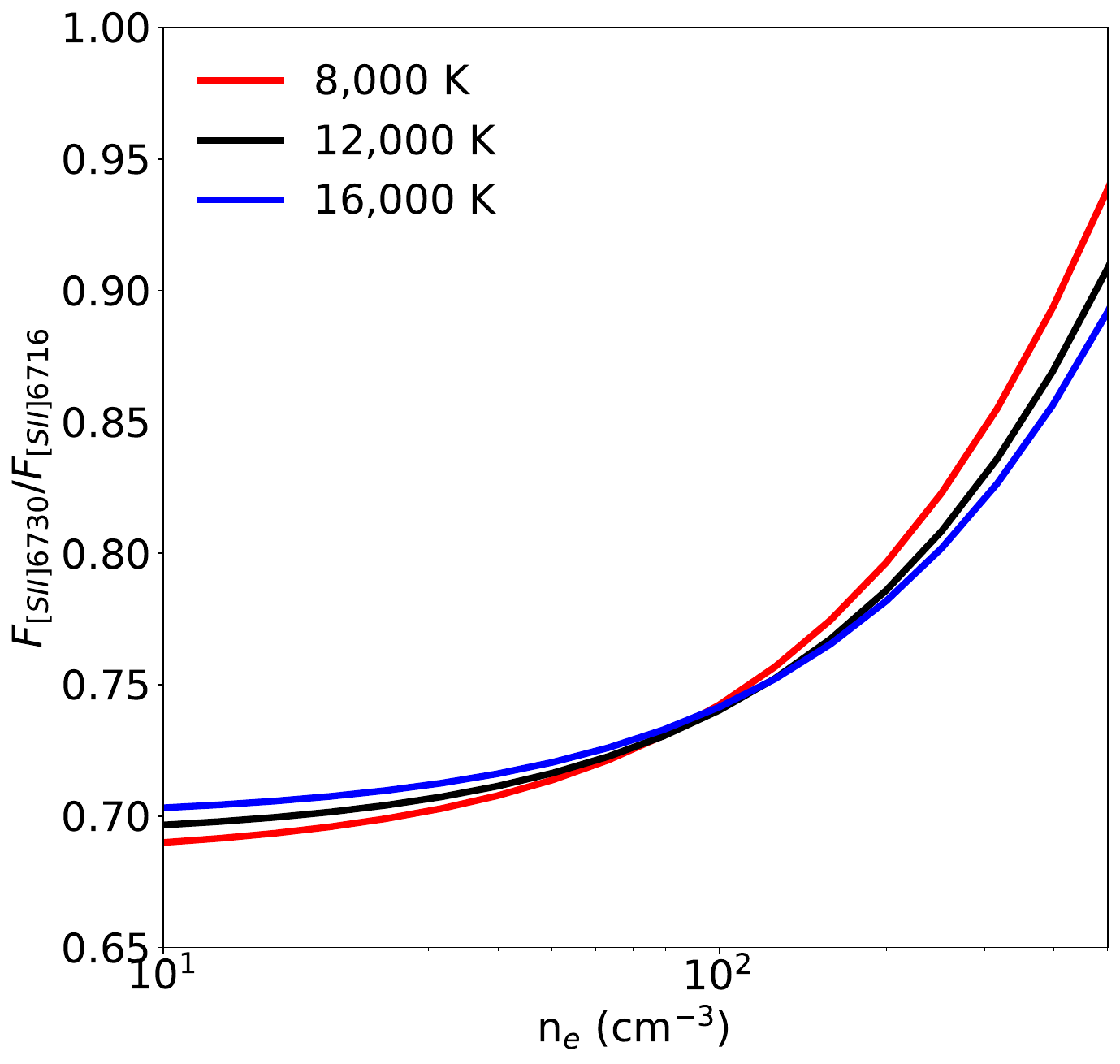}{(b)}
\caption{The density- and temperature-sensitive line-ratio diagnostics used to derive the physical conditions of the \HII\ regions within M101, plotted using PyNeb \citep{Luridiana2015}. The \oIII\,5007\AA/88\um\ ratio (a) is largely temperature sensitive, while the \sII\,6730\AA/6716\AA\ ratio (b) is density sensitive. We use both line-ratios simultaneously to constrain the electron temperature and density within the targeted \HII\ regions.}
\label{Theoretical_Diagnostic_Ratios}
\end{center}
\end{figure}

In this section, we use PyNeb \citep{Luridiana2015}, to simultaneously constrain the electron density and temperature within the targeted \HII\ regions of M101, using optical emission lines of ionized sulfur, \sII\,6730\AA\ and 6716\AA, whose ratio is density sensitive (see Figure \ref{Theoretical_Diagnostic_Ratios}a), as well as the FS transition of doubly-ionized oxygen, \oIII\,88\um\ in combination with the 5007\AA\ line, whose ratio is temperature sensitive (see Figure \ref{Theoretical_Diagnostic_Ratios}b). Pyneb uses the \oIII\ radiative transition probabilities of \cite{FroeseFischer2004} and collisional strengths of \cite{Storey2014}, exactly as does the CHAOS collaboration. These consistent atomic data allow us to directly compare the electron temperatures derived here on 20$\arcsec$ (520\,pc) spatial scales to those determined as part of the CHAOS program, which employs the \oIII\,4363\AA\ auroral line and higher spatial-resolution slit-spectroscopy.

For the \sII\,6730\AA/6716\AA\ line ratio from the PPAK observations, we apply only a 5\% absolute calibration uncertainty to each line, rather than the 10\% quoted in Table \ref{Flux_Table}. This smaller uncertainty is warranted because the two spectral lines are observed by the same instrument and have negligible differential extinction (the line-flux values presented here are always extinction-corrected), such that the absolute calibration is less important for the lines constituting this ratio. This assumption does not affect the results presented here. For the \oIII\,5007\AA/88\um\ ratio, we retain a 10\% absolute calibration uncertainty in each line, quoted in Table \ref{Flux_Table}, since the two lines are observed with different instruments and have different levels of dust attenuation.

The results of the simultaneous density and temperature determinations can be seen in Figure \ref{Region_Diagnostic_Plots_Last} and Table \ref{Region_Parameters_Table}. We note that the targeted \HII\ regions of M101 are always in, or very near, the low-density limit for the \sII\ line-pair, with measured $n_e$ \textless 300\,cm$^{-3}$ in all regions. This result is supported by observations of the \sIII\,18.7\um\ and 33.5\um\ lines, which also indicate low-density ($\lesssim$ 200 cm$^{-3}$) gas in the \HII\ regions of M101 \citep{Gordon2008}. This density regime is ideal for the FIR-abundance determination that we employ here, in that the emissivity-per-unit-volume of the \oIII\,88\um\ line, $\epsilon_{[O\,\textsc{iii}]88}$, is only mildly density-sensitive in this regime.

\begin{figure*}
\begin{center}		
\includegraphics[width=0.80\textwidth]{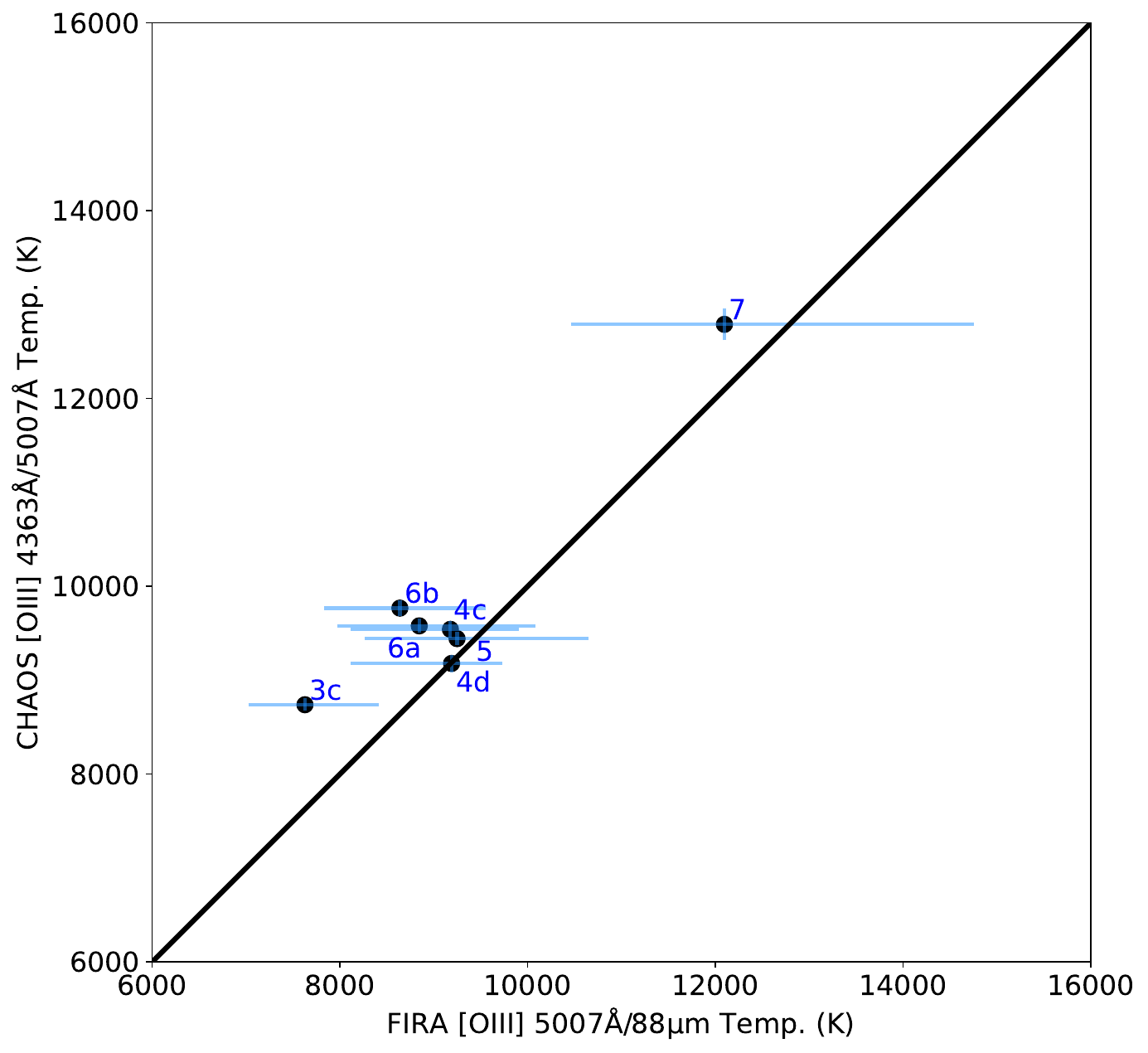}
\caption{Comparison between the FIRA-derived \HII\ region \oIII\ electron-temperatures, calculated using the \oIII\,5007\AA/88\um\ line-ratio, and those derived using the \oIII\,4363\AA/5007\AA\ auroral-line ratio \citep[CHAOS,][]{Croxall2016}. The two line-pairs produce temperatures that are quite consistent, with a mean difference of $\Delta T_e$ (CHAOS$-$FIRA) = 600 $\pm$ 400 K.}
\label{Temp_Comparison}
\end{center}
\end{figure*}

Using the \oIII\,5007\AA/88\um\ ratio, we measure electron temperatures over the range of $\sim$ 4,000$-$12,000\,K among all of our targeted regions. In Figure \ref{Temp_Comparison}, we compare these temperatures, measured on $\sim$ 0.5 kpc scales, to those derived in the $\sim$ 1\arcsec\ $\times$ 9\arcsec\ extraction slits of the CHAOS program, wherever the \oIII\,4363\AA\ auroral-line was detected. Overall, we see excellent agreement, with a mean difference of $\Delta T_e$ = 600\,K and a standard deviation of $\sigma T_e$ = 400\,K. Where a modest offset between the FIR/optical and auroral/optical temperature determinations may exist, the auroral/optical line method always produces the larger temperature. This discrepancy may be because the 4363\AA\ auroral-line has a higher excitation temperature than do either the FIR or optical \oIII\ fine-structure lines, such that it is preferentially emitted from regions of higher-temperature gas, leading to the higher measured temperature \citep[e.g., t$^2$;][]{Peimbert1967}. Or it could be that the \oIII\ 4363\AA\ line is contaminated by the nearby \feII\ 4360\AA\ line \citep[e.g.,][]{Curti2017}, although it appears that this contamination is only
important above 12 + log(O/H) $=$ 8.4. Alternatively, it could simply be that the discrepancies, seen only at the level of \textless\,2$\sigma$, are statistical in nature. A detailed comparison of these two temperature-diagnostics will be a major focus of the FIRA project.

The overall variation in temperature between the ionized-gas regions studied here makes clear the need for temperature diagnostics when using temperature-sensitive  (e.g., optical) methods to determine metal abundances. The excellent agreement between the temperatures determined with the FIR/optical and auroral/optical methods (with an average offset of $\Delta T_e$ = 600 $\pm$ 400 K), on scales of $\sim$ 20\arcsec\ (520 pc) and 1\arcsec\ (26 pc), respectively, suggests that if temperature variations are present in the gas, they are below our average temperature-uncertainty per \HII\ region ($\lesssim$ 1,000\,K). These results also support the validity of the \oIII\,5007\AA/88\um\ line ratio as an ionized-gas temperature-diagnostic, with the benefit that this line-pair is detected in regions down to substantially lower temperatures, and fainter \HII\ regions, than is possible with the \oIII\,4363\AA\ auroral line (e.g., Nuc and Region 1). The difficulty in detecting the \oIII\,4363\AA\ line at small galacto-centric radii (high metallicity) is two-fold -- at high metallicity, electron-temperature decreases and doubly-ionized oxygen becomes the sub-dominant species to singly-ionized oxygen -- while the FIR \oIII\ line emission is only affected by the decreasing fraction of O$^{++}$ ions, not by the decreasing gas-temperature.

\begin{figure*}
\begin{center}
\begin{tabular}{cc}
\includegraphics[width=0.45\textwidth]{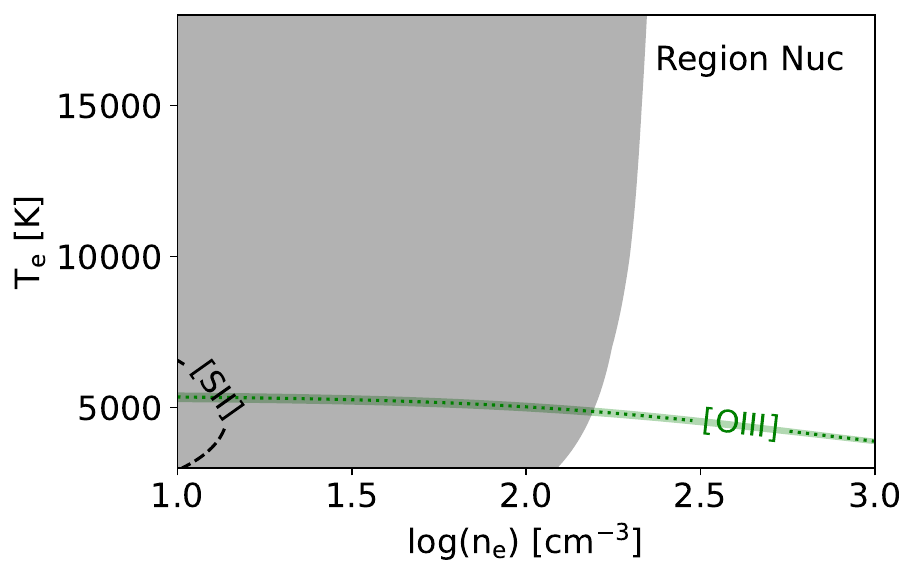} &
\includegraphics[width=0.45\textwidth]{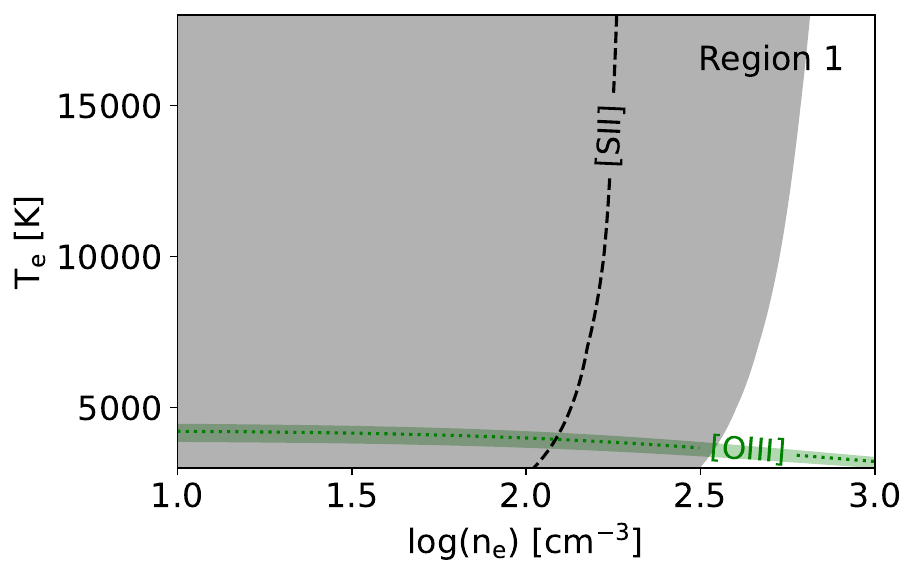} \\
\includegraphics[width=0.45\textwidth]{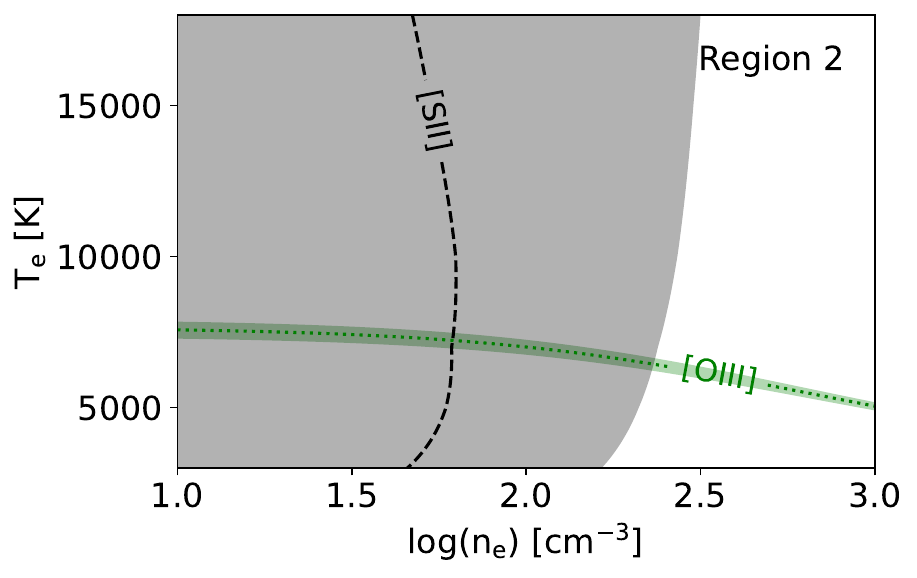} &
\includegraphics[width=0.45\textwidth]{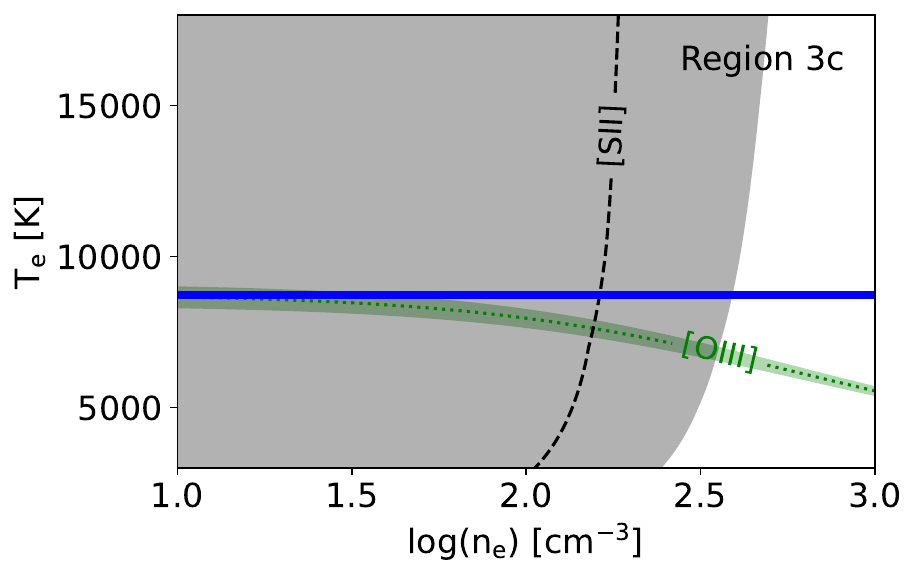} \\
\includegraphics[width=0.45\textwidth]{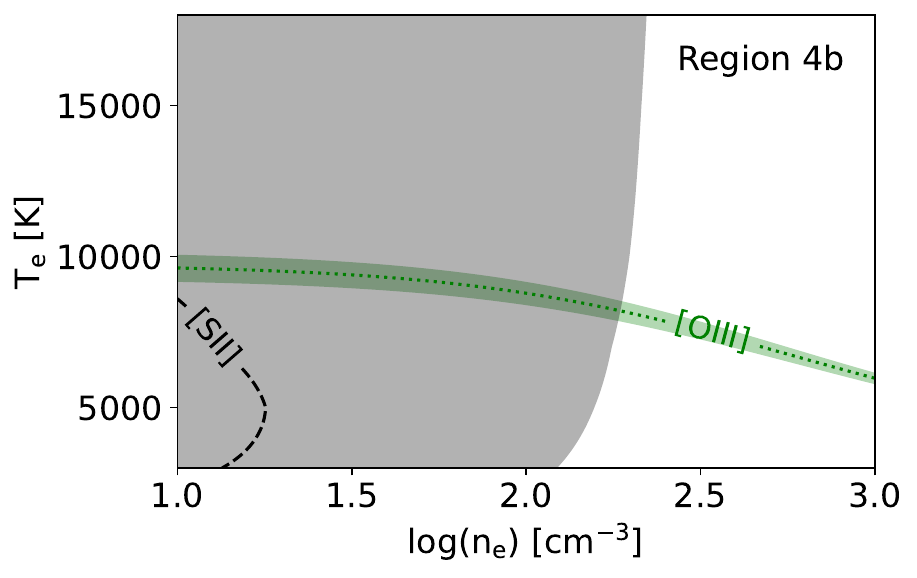} &
\includegraphics[width=0.45\textwidth]{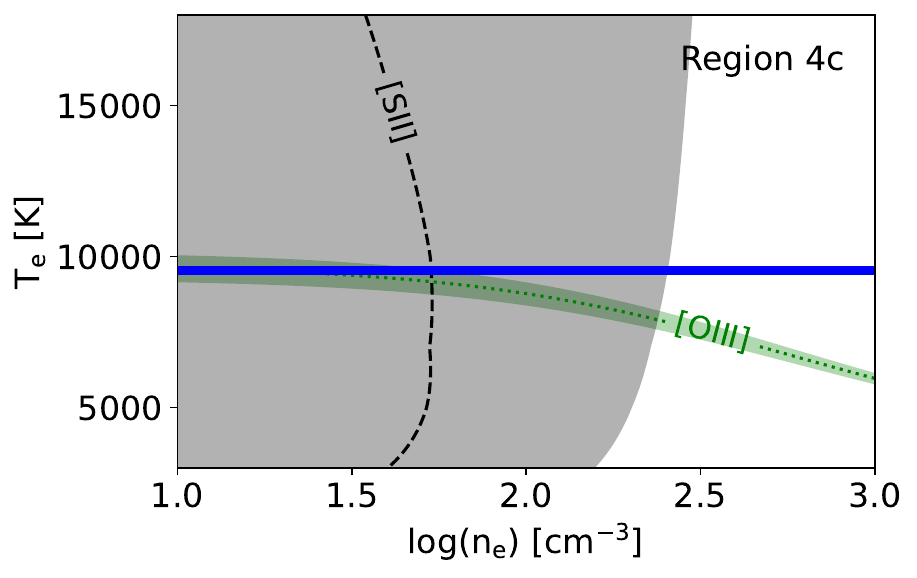} \\
\end{tabular}
\end{center}
\end{figure*}

\begin{figure*}
\begin{center}
\begin{tabular}{cc}
\includegraphics[width=0.45\textwidth]{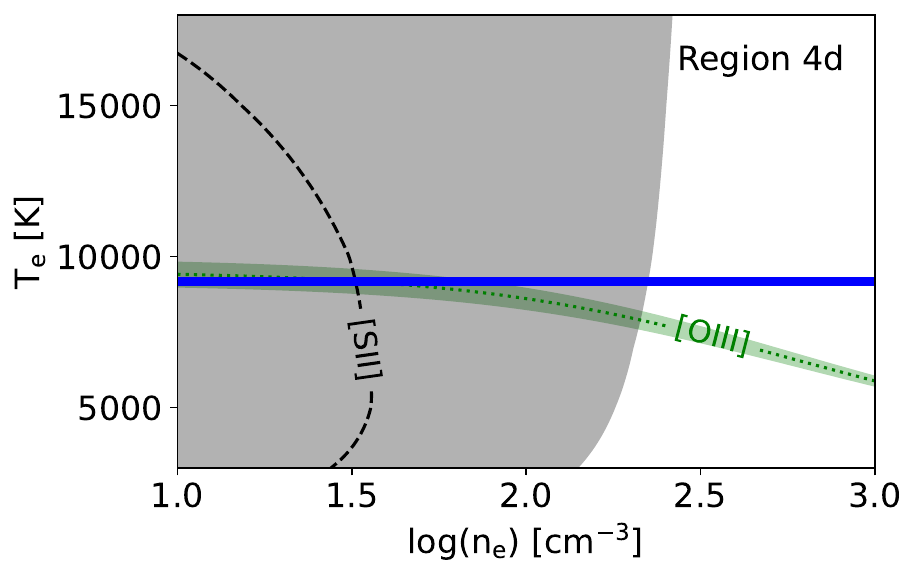} &
\includegraphics[width=0.45\textwidth]{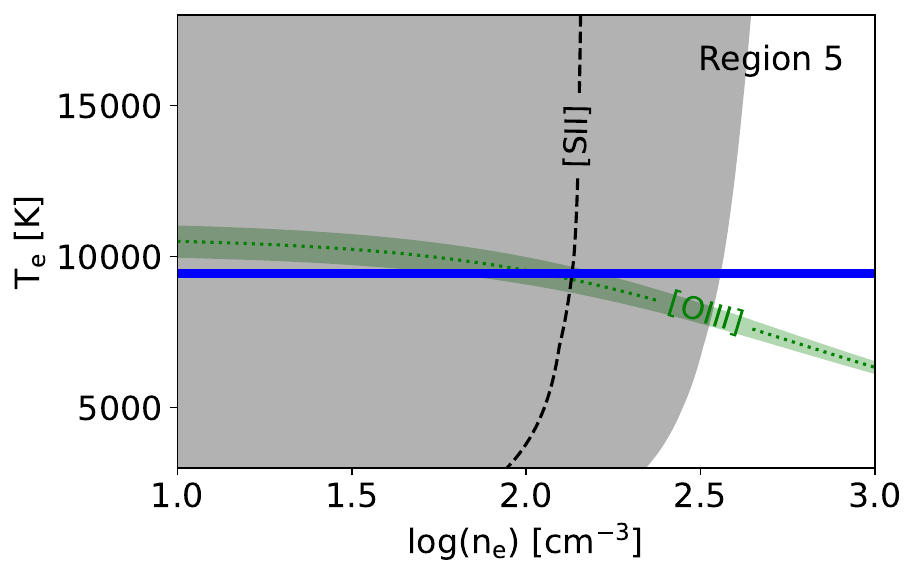} \\
\includegraphics[width=0.45\textwidth]{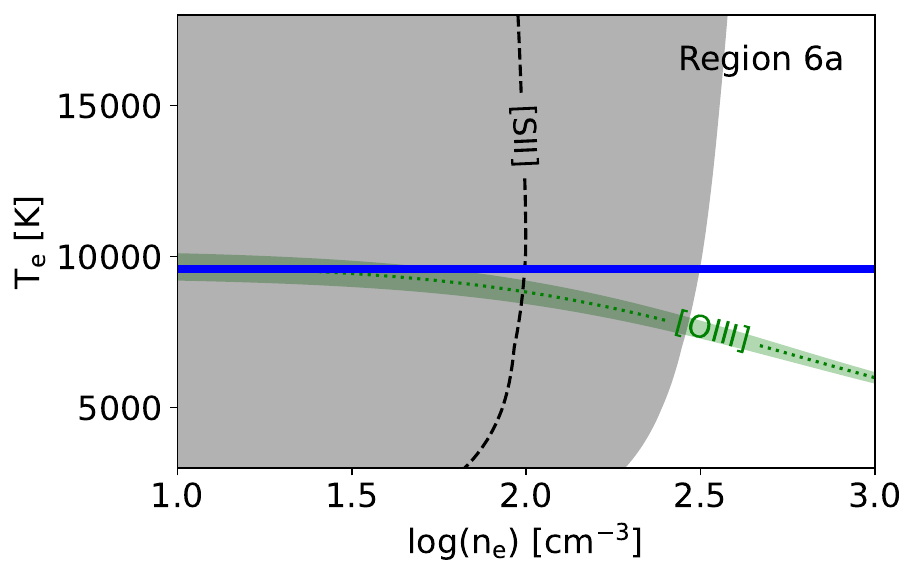} &
\includegraphics[width=0.45\textwidth]{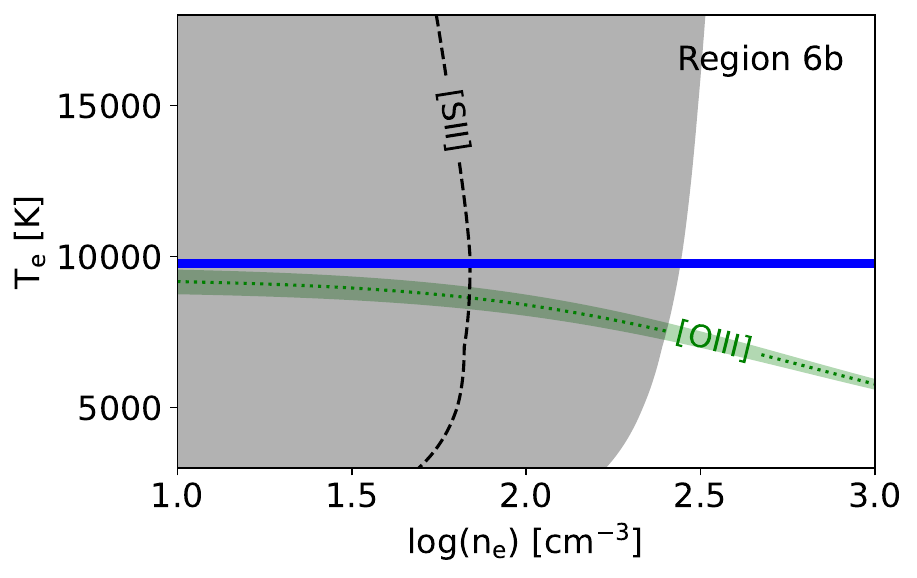} \\
\includegraphics[width=0.45\textwidth]{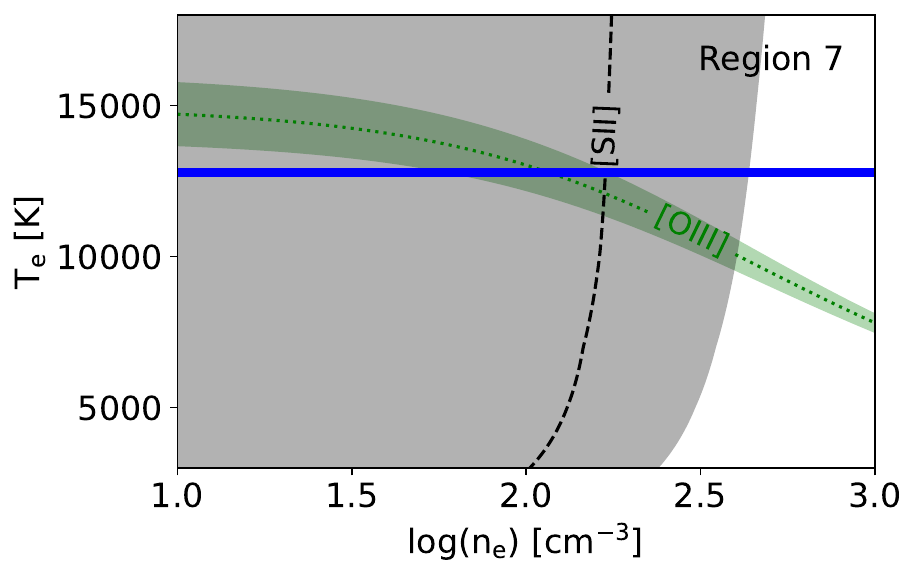} 
\end{tabular}
\caption{(Continued from previous page) Derivation of the physical parameters within the \HII\ regions of M101. Grey (green) shading indicates parameter space allowed by the measured \sII\,6730\AA/6716\AA\ density-sensitive (\oIII\,5007\AA/88\um\ temperature-sensitive) line-ratio, such that the doubly-shaded regions are allowed by both. The horizontal blue lines, where present, indicate the [OIII] auroral-line-derived temperatures from CHAOS \citep{Croxall2016}. Figures created using PyNeb \citep{Luridiana2015}.}
\label{Region_Diagnostic_Plots_Last}
\end{center}
\end{figure*}

\begin{deluxetable}{ccccc}
\tablecaption{Derived ISM Parameters}
\tablecolumns{5}
\tablenum{3}
\tablehead{
\colhead{Region} & \colhead{T$_e$ (FIRA)} & \colhead{n$_e$ (FIRA)} & \colhead{T$_e$ (CHAOS)\tablenotemark{a}} & \colhead{n$_e$ (CHAOS)\tablenotemark{a}}\\
\colhead{} & \colhead{(K)} & \colhead{(cm$^{-3}$)} & \colhead{(K)} & \colhead{(cm$^{-3}$)} \\
}
\startdata
Nuc & $5340 ^{+ 110 }_{- 360 }$ & \textless 90 & ... & ... \\
1 & $3950 ^{+ 430 }_{- 490 }$ & \textless 300 & ... & ... \\
2 & $7220 ^{+ 610 }_{- 570 }$ & \textless 140 & ... & 100 $\pm$ 100 \\
3a & ... & ... & 8580 $\pm$ 140 & 100 $\pm$ 100 \\
3b & ... & ... & 9170 $\pm$ 190 & 100 $\pm$ 100 \\
3c & $7630 ^{+ 790 }_{- 600 }$ & \textless 250 & 8739 $\pm$ 66 & 272 $\pm$ 137 \\
3d & ... & ... & ... & ... \\
4a & ... & ... & ... & ... \\
4b & $9660 ^{+ 250 }_{- 1090 }$ & \textless 100 & ... & ... \\
4c & $9180 ^{+ 730 }_{- 1060 }$ & \textless 140 & 9542 $\pm$ 90 & 100 $\pm$ 100 \\
4d & $9190 ^{+ 540 }_{- 1070 }$ & \textless 120 & 9179 $\pm$ 93 & 100 $\pm$ 100 \\
5 & $9250 ^{+ 1400 }_{- 980 }$ & \textless 240 & 9443 $\pm$ 70 & 235 $\pm$ 100 \\
6a & $8840 ^{+ 1240 }_{- 870 }$ & \textless 190 & 9579 $\pm$ 75 & 130 $\pm$ 74 \\
6b & $8640 ^{+ 920 }_{- 810 }$ & \textless 160 & 9768 $\pm$ 81 & 100 $\pm$ 100 \\
6c & ... & ... & 9299 $\pm$ 69 & 100 $\pm$ 67 \\
7 & $12100 ^{+ 2650 }_{- 1640 }$ & \textless 280 & 12790 $\pm$ 170 & 248 $\pm$ 100 \\
\enddata
\tablecomments{The derived \oIII\ electron-temperatures and densities within the targeted \HII\ regions of M101. The FIRA \oIII\ electron-temperatures (presented here) are calculated using the \oIII\,5007\AA/88\um\ line-ratio, while the CHAOS \oIII\ electron-temperatures are calculated using the \oIII\,4363\AA/5007\AA\ line-ratio. For both FIRA and CHAOS, electron densities are calculated using the \sII\,6730\AA/6716\AA\ line-ratio.}
\tablenotetext{a}{\cite{Croxall2016}}
\label{Region_Parameters_Table}
\end{deluxetable}

\section{Choice Of Hydrogen Normalization}

For any determination of absolute gas-phase abundances, some scheme for hydrogen normalization must be employed. Here, with both high-frequency radio-continuum maps, which trace free-free emission, and H$\alpha$ recombination maps, we can compare two different methods for determining the hydrogen normalization that should, at least in theory, produce similar results. A systematic comparison of these two hydrogen normalizations is a major focus of the FIRA project, and we introduce the comparison here.

First, we note that not all radio emission, even at high-frequency, is the result of free-free interactions. The SFRS survey \citep{Linden2020}, conducted with the VLA, has obtained not only 33 GHz continuum imaging, but also lower-frequency imaging, at 15 and 3 GHz. This allows for the decomposition of the radio continuum into thermal free-free and synchrotron components, by fitting two power-laws. The thermal-emission power-law index is fixed at -0.1, while the non-thermal power-law index is allowed to vary. We adopt the derived free-free fraction at 33GHz, calculated as part of the SFRS survey, in each of the targeted \HII\ regions of M101, using that fraction to convert from observed 33GHz-flux to free-free flux in our apertures (see Table \ref{Flux_Table}). 

We note that the free-free fraction at 33GHz is always very high, $\sim$ 80$-$100\%, in the \HII\ regions of M101, and indeed the SFRS survey finds a median free-free fraction at 33GHz of $\sim$ 93\% in their entire sample of extra-galactic \HII\ regions. This is fortunate, as multi-band high-resolution radio-continuum observations may not be available for every target of interest for the application of the FIR-abundances method. In these cases without multi-band radio observations, a single high-frequency continuum-image can be obtained, with the free-free emission assumed to be a large fraction of the total observed flux \citep[e.g., $\sim$ 93\%,][]{Linden2020}. This assumption breaks down for galaxies that contain AGN, if the spatial resolution of the radio observations is not sufficiently high to disentangle the targeted \HII\ regions from the central emission (e.g., high-redshift galaxies), and in ionized-gas regions dominated by emission from supernovae remnants.

With the thermal free-free flux estimated, and the temperature and density of the \HII\ regions derived above, the theoretical $F_{\rm H \alpha}$/$S_{\rm ff, \nu}$ ratio is given by the following equation \citep[where we adopt the free-free emissivity from][]{DraineBook}:

\begin{equation}
\frac{F_{\rm H \alpha} ({\rm erg\,s^{-1}\,cm^{-2})}}{S_{ff, \nu} ({\rm Jy})} = \frac{\epsilon_{H \alpha}(n_e,T_e) ({\rm erg\,s^{-1}\,cm^{3}})}{4.21 \times 10^{-16} \cdot T_{4}^{-0.323} \cdot \nu_{9}^{-0.118}},
\end{equation}

\noindent where $F_{\rm H \alpha}$ is the H$\alpha$ line flux, $S_{\rm \nu,ff}$ is the free-free contribution to the flux-density at frequency $\nu$, $T_4$ is the \HII-region electron-temperature in units of 10$^4$ K, $\nu_9$ is the radio-continuum frequency in units of GHz, $\epsilon_{\rm H \alpha}(n_e,T_e)$ is the emissivity per unit volume of the H$\alpha$ line, at density $n_e$ and temperature $T_e$. Note that, at reasonable ISM densities, this ratio is largely insensitive to density, though it is slightly temperature dependent ($\propto$ $T_e^{-0.5}$).

\begin{figure*}
\begin{center}		
\includegraphics[width=0.95\textwidth]{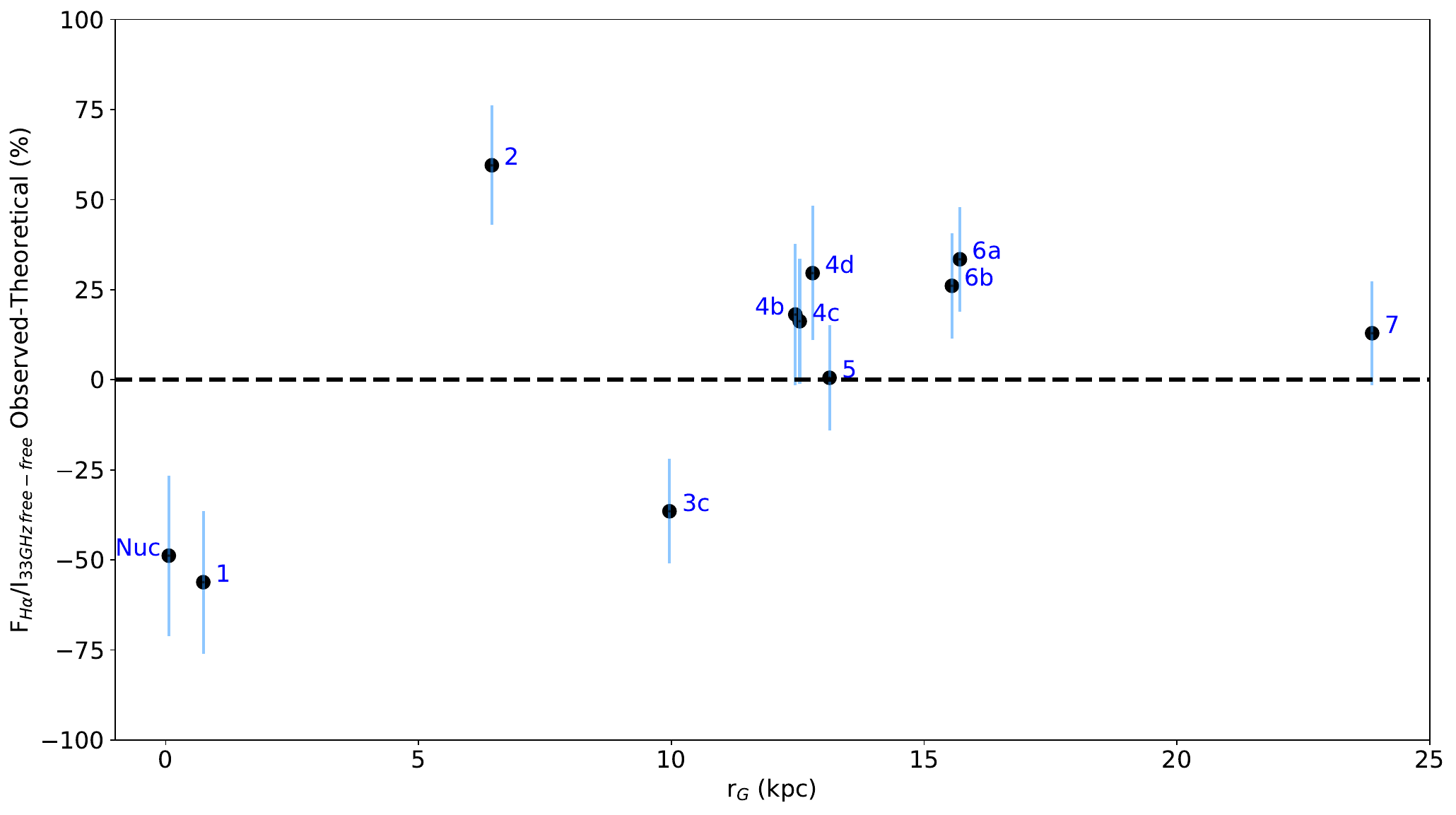}
\caption{Percentage difference between the observed and theoretical $F_{\rm H \alpha}$/$S_{\rm ff, 33GHz}$ ratio, calculated at the derived densities and temperatures for each \HII\ region, as a function of galacto-centric radius. Since the theoretical ratio is calculated at the derived density and temperature for each \HII\ region individually, the dashed y=0 line would indicate prefect agreement between the expected and measured values. The observed $F_{\rm H \alpha}$/$S_{\rm ff, 33GHz}$ ratios scatters around the expected values, with a fractional scatter of $\sim$\,39\%. This scatter may be due to several factors, including H$\alpha$ contribution from diffuse ionized-gas, errors in the applied optical attenuation-correction factors, or resolving out the free-free continuum by the VLA interferometer.}
\label{Theoretical_Measured_Ha_ff}
\end{center}
\end{figure*}

In Figure \ref{Theoretical_Measured_Ha_ff}, we plot the difference between the observed (using 20\arcsec\ extraction-apertures) and theoretical $F_{\rm H \alpha}$/$S_{\rm ff, 33GHz}$ ratios, calculated at the derived densities and temperatures for each \HII\ region individually, as a function of galactic radius.  The measured $F_{\rm H \alpha}$/$S_{\rm ff, 33GHz}$ ratios scatter around the expected values for the \HII\ regions of M101, within a fractional scatter of $\sim$\,39\%. We further find that the measured ratio within a given \HII\ region varies as a function of extraction aperture, typically increasing by $\sim$ 5$-$10\% between 10\arcsec\ to 20\arcsec\ (with both maps convolved to a common resolution of 10\arcsec). This ratio should be constant with extraction aperture, if both the radio and H$\alpha$ emission originate from the same gas. The increasing ratio with extraction aperture could be attributed to a spatially-varying error in the optical attenuation-correction, or a resolving-out of the interferometric radio-continuum emission on larger scales. Unfortunately, the 12.37\um\ Humphreys-$\alpha$ line, which would reveal any error in the adopted optical attenuation-correction factors, is not detected in the high-res Spitzer/IRS observations of M101. This non-detection is consistent with the expected Humphreys-$\alpha$ line-flux, estimated from the measured H$\alpha$ line-flux, and sensitivity of the Spitzer/IRS observations.

A change in the fraction of radio flux attributed to free-free emission could also cause variation in the $F_{\rm H \alpha}$/$S_{\rm ff, 33GHz}$ ratio. However, to flatten the increasing ratio with extraction aperture seen here would require increasing the free-free fraction at larger radii, in opposition to the expected trend that larger-scale radio-emission is usually synchrotron dominated. Indeed, \cite{Murphy2012} found that the median thermal free-free fraction at 33\,GHz, measured in a sample of extra-galactic \HII\ regions using the single-dish Green Bank Telescope (GBT), which has a beam FWHM of $\sim$25\arcsec\ at 33\,GHz, was $\sim$ 80\%. They further found that \HII\ regions resolved by the GBT beam had higher thermal fractions ($\sim$90\%), indicating that larger-scale emission, which fills a larger fraction of the beam for the unresolved sources, is increasingly non-thermal. Thus, we do not attribute this changing $F_{\rm H \alpha}$/$S_{\rm ff, 33GHz}$ ratio to changes in the makeup of the radio SED.

In the following analysis, we calculate the absolute gas-phase oxygen-abundance using both hydrogen normalizations separately, comparing the difference between the two.

\section{FIR-Derived Ionic Abundances}

The use of FIR fine-structure lines, together with radio free-free emission, to determine the gas-phase abundances of \HII\ regions has been employed in the Milky Way \citep[e.g.,][]{Herter1981, Rudolph1997} and also in high-redshift galaxies \citep[e.g.,][]{Lamarche2018}. FIR-FS lines have also been used in conjunction with optical hydrogen-recombination lines to derive abundances in local galaxies \citep[e.g.,][]{Croxall2013}. Here we employ both of these techniques in M101, where we have excellent overlapping observations with direct-abundance measurements from CHAOS \citep[][]{Berg2015, Croxall2016}. We note that the CHAOS group has recently published updated gas-phase abundances in M101 \citep{Berg2020}, using a multi-zone temperature model; however, we compare with the results of \cite{Croxall2016}, which used the derived \oIII\ temperatures to calculate the abundances, as we do here, for a more direct comparison \citep[and see also][in which the authors also employ optical direct-abundance techniques in M101 and find an oxygen abundance-gradient consistent with that found in \cite{Croxall2016}]{Esteban2020}.

The gas-phase O$^{++}$/H$^+$ abundance can be calculated using the \oIII\,88\um\ line together with radio free-free emission, from the ratio of fluxes:

\begin{equation}
\frac{n_{\rm O^{++}}}{n_{\rm H^+}} = \frac{F_{\rm [OIII]88}}{S_{\rm 33GHz,ff}} \frac{4.21 \times 10^{-16} \cdot T_{4}^{-0.323} \cdot \nu_{9}^{-0.118}}{\epsilon_{\rm [OIII]88}(T_e, n_e)} \left( \frac{n_e}{n_p} \right),
\end{equation}

\noindent where $n_{\rm O^{++}}$/$n_{\rm H^+}$ is the abundance of O$^{++}$ relative to hydrogen, $F_{\rm \oIII\,88\um}$ is the \oIII\,88\um\ line flux in units of ergs s$^{-1}$ cm$^{-2}$, $\epsilon_{\rm [O\,\textsc{iii}]88}(T_e,n_e)$ is the emissivity per unit volume of the \oIII\,88\um\ line, in units of ergs s$^{-1}$ cm$^{3}$, and $n_e/n_p$ is the electron to proton number-density ratio, which accounts for the contribution of electrons from non-hydrogen atoms present in the H\,{\sc ii} regions. We adopt the density and temperature values calculated above for each \HII\ region individually and determine the \oIII\,88\um\ line emissivity using PyNeb \citep{Luridiana2015}. We also assume $n_e$/$n_p$ = 1.05, which accounts for the electrons contributed from helium, the second most abundant element. This ratio has mild temperature and density dependencies, at densities $\lesssim$ 200 cm$^{-3}$, varying by $\sim$ 10\% over the range $T = $ 5,000 -- 12,000\,K, and by a factor of two over the range $n = $ 10 -- 200 cm$^{-3}$.

Equivalently, the gas-phase O$^{++}$/H$^+$ abundance can be calculated using the H$\alpha$ recombination line, from the equation:

\begin{equation}
\frac{n_{\rm O^{++}}}{n_{H^+}} = \frac{F_{\rm [OIII]88}}{F_{\rm H \alpha}} \frac{\epsilon_{\rm H \alpha}(n,T)}{\epsilon_{\rm [OIII]88}(n, T)},
\end{equation}

\noindent where the variables are as in the previous equation. And similarly, this ratio has mild temperature and density dependencies, at densities $\lesssim$ 200 cm$^{-3}$, varying by a factor of two over the range $T = $ 5,000 -- 12,000\,K, and by a factor of two over the range $n = $ 10 -- 200 cm$^{-3}$.

\begin{deluxetable*}{cccc}
\tablecaption{Derived Ionic Abundances}
\tablecolumns{4}
\tablenum{4}
\tablehead{
\colhead{Region} & \colhead{O$^{++}$/H$^{+}$ (FIR/H$\alpha$)} & \colhead{O$^{++}$/H$^{+}$ (FIR/Free-Free)} & \colhead{O$^{++}$/H$^{+}$ (Optical Direct-Abundance)\tablenotemark{a}} \\
\colhead{} & \colhead{(10$^{-4}$)} & \colhead{(10$^{-4}$)} & \colhead{(10$^{-4}$)} \\
}
\startdata
Nuc & $ 0.55 ^{+ 0.23 }_{- 0.08 }$ & $0.30 ^{+ 0.12 }_{- 0.07 }$ & ... \\
1 & $ 1.66 ^{+ 1.35 }_{- 0.69 }$ & $0.76 ^{+ 0.54 }_{- 0.31 }$ & ... \\
2 & $ 0.27 ^{+ 0.10 }_{- 0.07 }$ & $0.45 ^{+ 0.15 }_{- 0.11 }$ & 0.43 \,$\pm$\, 0.04 \\
3a & ... & ... & 1.10 \,$\pm$\, 0.04 \\
3b & ... & ... & 0.79 \,$\pm$\, 0.03 \\
3c & $ 2.88 ^{+ 1.02 }_{- 0.92 }$ & $1.92 ^{+ 0.58 }_{- 0.54 }$ & 1.97 \,$\pm$\, 0.04 \\
3d & ... & ... & ... \\
4a & ... & ... & ... \\
4b & $ 1.04 ^{+ 0.47 }_{- 0.15 }$ & $1.29 ^{+ 0.50 }_{- 0.25 }$ & ... \\
4c & $ 1.25 ^{+ 0.53 }_{- 0.28 }$ & $1.52 ^{+ 0.53 }_{- 0.33 }$ & 1.45 \,$\pm$\, 0.03 \\
4d & $ 1.21 ^{+ 0.54 }_{- 0.21 }$ & $1.65 ^{+ 0.61 }_{- 0.33 }$ & 1.70 \,$\pm$\, 0.04 \\
5 & $ 1.49 ^{+ 0.61 }_{- 0.52 }$ & $1.57 ^{+ 0.51 }_{- 0.47 }$ & 1.69 \,$\pm$\, 0.03 \\
6a & $ 1.62 ^{+ 0.63 }_{- 0.53 }$ & $2.27 ^{+ 0.71 }_{- 0.64 }$ & 1.55 \,$\pm$\, 0.03 \\
6b & $ 1.56 ^{+ 0.61 }_{- 0.41 }$ & $2.06 ^{+ 0.66 }_{- 0.47 }$ & 1.03 \,$\pm$\, 0.02 \\
6c & ... & ... & 1.50 \,$\pm$\, 0.03 \\
7 & $ 1.11 ^{+ 0.50 }_{- 0.43 }$ & $1.31 ^{+ 0.44 }_{- 0.41 }$ & 1.07 \,$\pm$\, 0.02 \\
\enddata
\tablecomments{The derived O$^{++}$/H$^{+}$ ionic abundances, calculated by FIRA using the \oIII\,88\um\ line and either H$\alpha$ or free-free emission for the hydrogen normalization, compared to those calculated by the CHAOS collaboration using collisionally-excited optical direct-abundance methods.}
\tablenotetext{a}{\cite{Croxall2016}}
\label{Ionic_Abundances_Table}
\end{deluxetable*}

\begin{figure*}
\begin{center}		
\begin{tabular}{cc}
\includegraphics[width=0.65\textwidth]{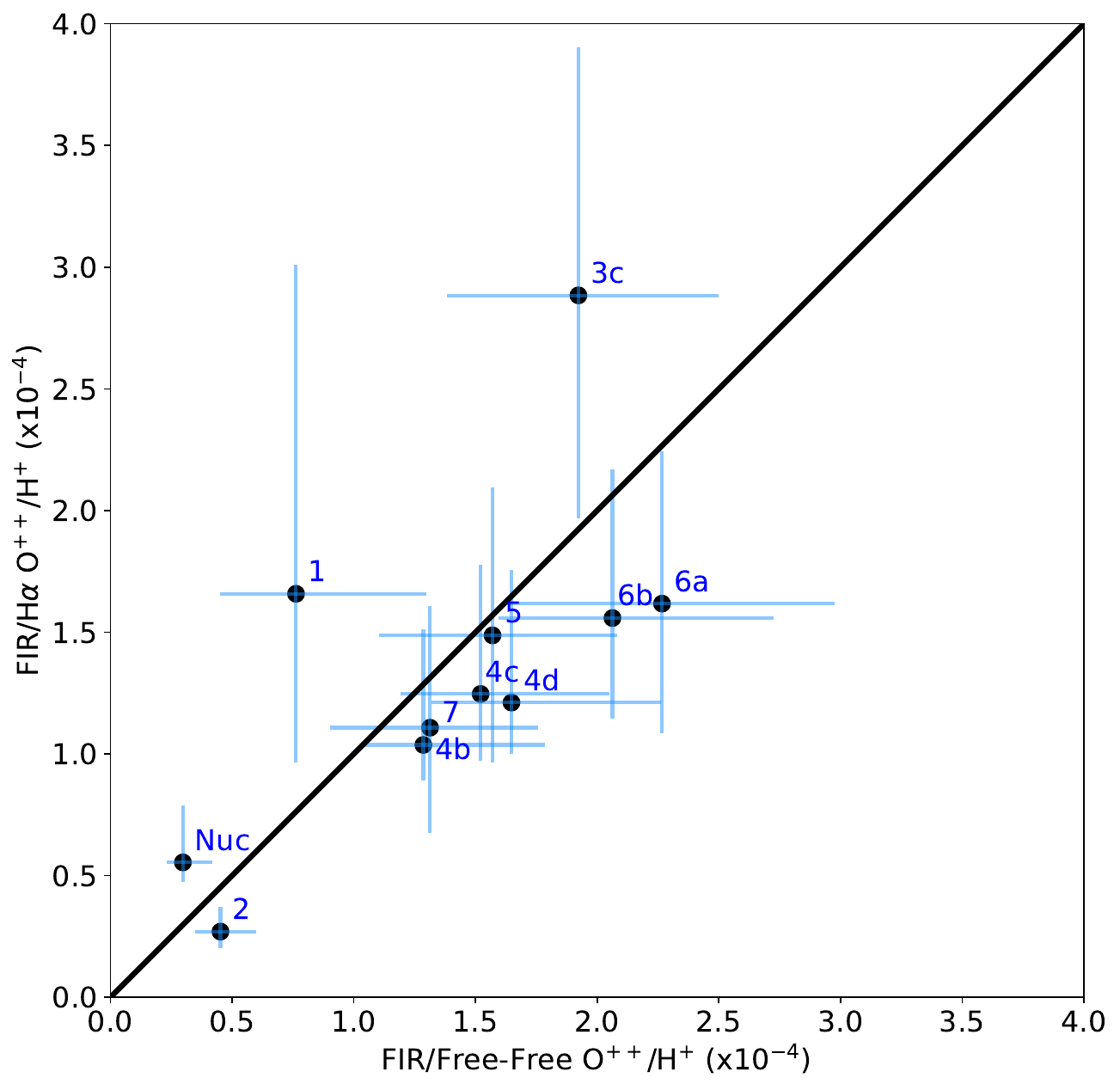}{(a)} \\
\includegraphics[width=0.45\textwidth]{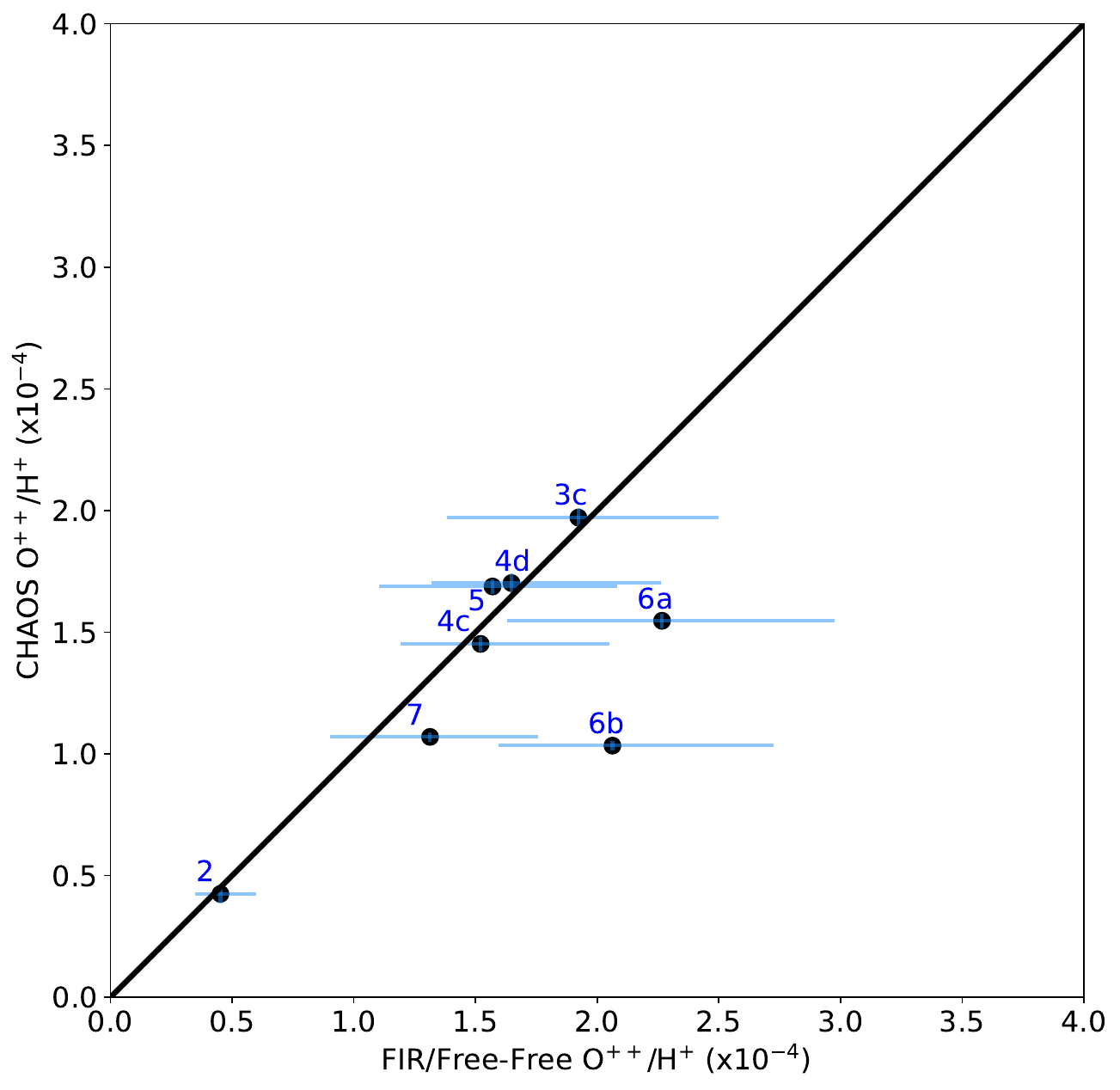}{(b)}
\includegraphics[width=0.45\textwidth]{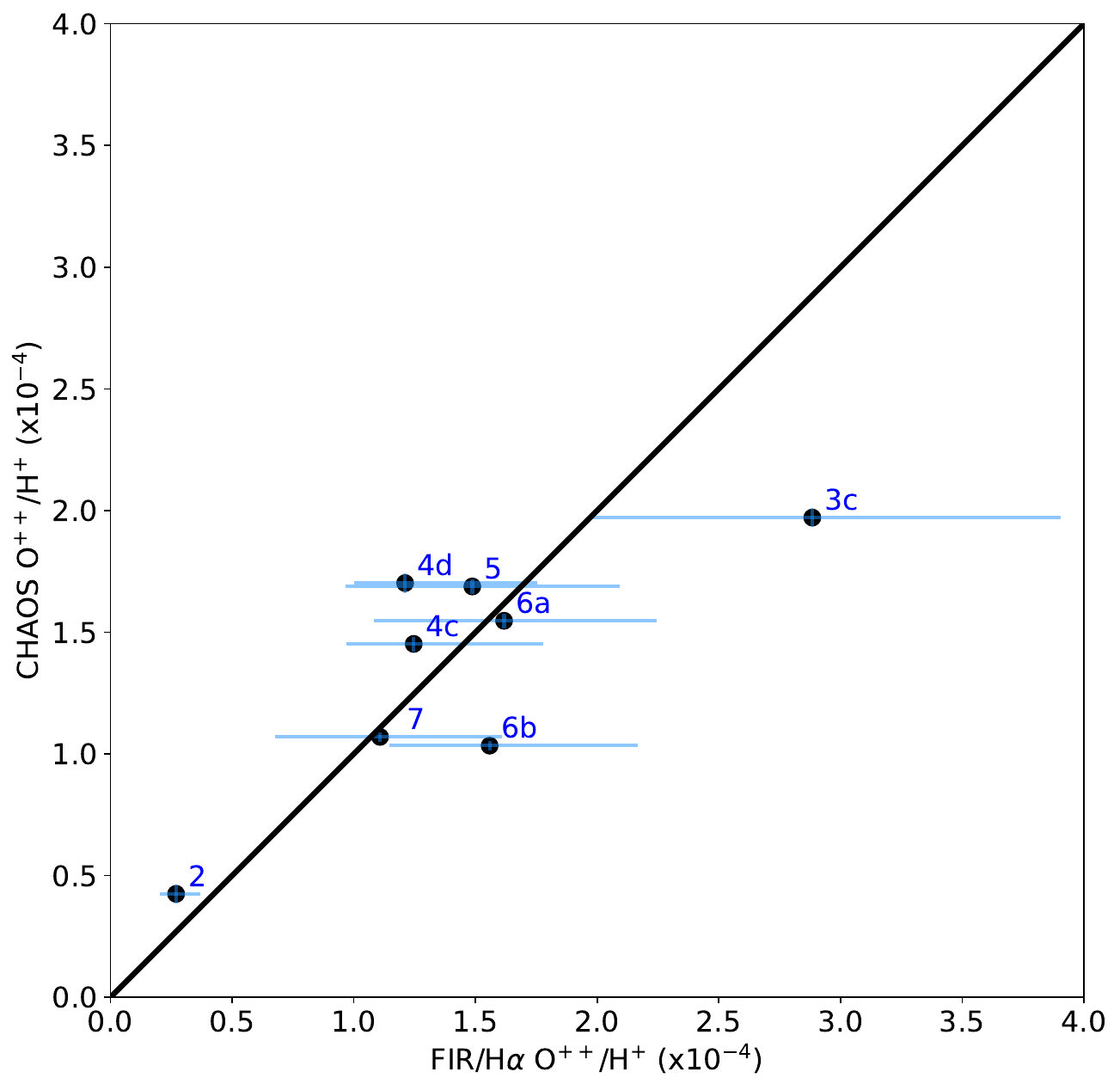}{(c)}
\end{tabular}
\caption{Comparison between the O$^{++}$/H$^+$ ionic abundances measured by FIRA (this work) and CHAOS temperature-corrected auroral-line methods \citep[][]{Croxall2016}. (a) O$^{++}$/H$^+$ ionic abundances derived using the FIR/free-free and FIR/H$\alpha$ normalizations are in good agreement, within the fractional scatter of 39\%, due to the scatter in the hydrogen normalizations. Similarly, both the FIR/free-free (b) and FIR/H$\alpha$ (c) normalizations are in good agreement with CHAOS, within the fractional scatters of 24\% and 29\%, respectively. \HII\ regions appearing in (a) but absent from (b) and (c) are not detected in the \oIII\,4363\AA\ line by the CHAOS program.}
\label{Ionic_Abundances_Figure}
\end{center}
\end{figure*}

The O$^{++}$ abundances derived using the FIR \oIII\,88\um\ line, in conjunction with radio free-free and H$\alpha$ emission, are in broad agreement with those derived from the CHAOS direct-abundance measurements. We find that the FIRA ionic-abundances calculated using both free-free and H$\alpha$ emission as the hydrogen normalizations are consistent with the values determined by CHAOS, within the fractional scatter of 24\% and 29\%, respectively (see Figure \ref{Ionic_Abundances_Figure} and Table \ref{Ionic_Abundances_Table}).

\section{Ionization Determination}

The ionization state of the targeted \HII\ regions in M101 must next be estimated if the ionic abundances determined above are to be converted into total elemental-abundances. Unfortunately, calculation of the so called ``ionization correction factor" (ICF) for the oxygen atom/ion in the far-infrared must necessarily be indirect. This is because the O$^{+}$ ion has no ground-state fine-structure splitting and hence no FIR emission-lines. This is not a significant concern in the optical, where the O$^{+}$ ion does have emission lines, and both O$^0$ and O$^{+++}$ are presumed to have very low abundances, provided that the electron temperatures can be measured.

To overcome this limitation, which is most significant at low O$^{++}$/O$^+$ relative abundances (generally at higher metallicity), we can employ an ICF-sensitive ionization-parameter indicator.  For this indicator, we adopt the ratio of the mid-IR \neIII\,15.6\um\ and \neII\,12.8\um\ lines from the Spitzer/IRS observations. We compare this observational hardness-indicator with directly-measured optical O$^{++}$/O$^+$ ratios from CHAOS, where available, as well as with \HII\ region photoionization models, to constrain the ICF. In the following discussion, we use the Mexican Million Models Database \citep[3MdB,][]{Morisset2015, ValeAsari2016}, which has been employed previously in determining the gas-phase abundances of ionized regions from FIR-FS line-measurements \citep[e.g.,][]{Peng2021}.

\begin{figure*}
\begin{center}		
\includegraphics[width=0.95\textwidth]{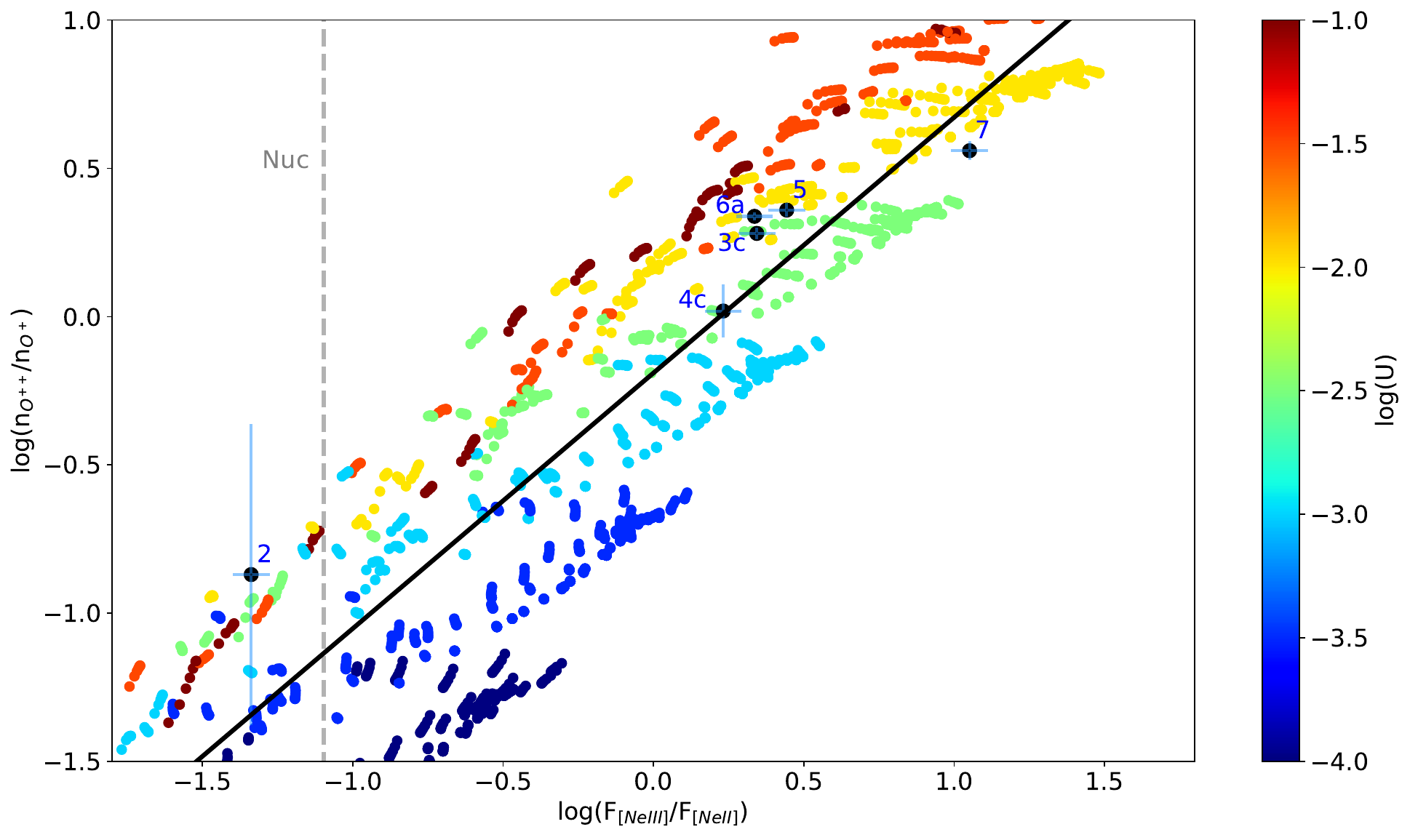}
\caption{Derivation of the n$_{\rm O^{++}}$/n$_{\rm O^{+}}$ abundance-ratio using the \neIII/\neII\ line-ratio as a proxy. The black solid-line indicates the best-fit to the photoionization-model data-points \citep[color, 3MdB;][]{Morisset2015, ValeAsari2016}, while the black points indicate measured values \citep[n$_{\rm O^{++}}$/n$_{\rm O^{+}}$ abundance-ratios from][Spitzer line-fluxes presented here]{Croxall2016}. The color-bar indicates the ionization parameter, log(U), for each model. Since the O$^+$ ion has no fine-structure transitions in the FIR, in \HII\ regions where the n$_{\rm O^{++}}$/n$_{\rm O^{+}}$ abundance-ratio cannot be directly measured in the optical, such as the nucleus (indicated by the vertical grey dashed line), we employ this best-fit to determine the ICF from measurements of the \neIII/\neII\ ratio in the mid-IR.}
\label{fig:icf}
\end{center}
\end{figure*}

The colored points in Figure~\ref{fig:icf} show the numerical models from the BOND 3MdB \citep{ValeAsari2016}. These photoionization models are run for an ensemble of \HII\ regions, using the modeling software \emph{Cloudy} \citep[e.g.,][]{Ferland2017}, version 17.02, with electron densities of $n_e$ = 100 cm$^{-3}$, a range in metallicity of 6.6 \textless\ 12+log(O/H) \textless\ 9.4, and a range in input ionization-parameter of -4 \textless\ log(U) \textless\ -1. The input SED was modeled as an ensemble stellar population, with ages ranging from 1--6 Myrs, determined using the population synthesis code PopStar \citep{Molla2009}, with a \cite{Chabrier2003} stellar initial-mass function. We consider only those models with a filled, spherical geometry, which are radiation bounded (H$\beta$ depth \textgreater\, 95\%). We further constrain the models by considering only those with electron temperatures in the range measured here (see Table \ref{Region_Parameters_Table}). The over-plotted black line in Fig.~\ref{fig:icf} shows the trend of the model relationship between the [Ne\,{\sc iii}]15.6\um/[Ne\,{\sc ii}]12.8\um\ ratio and the O$^{++}$/O$^+$ abundance-ratio. The best-fit relation is given by the equation:

\begin{equation}
\log_{10} \left( \frac{n_{\rm O^{++}}}{n_{\rm O^{+}}} \right) = 0.86\,\log_{10} \left( \frac{F_{\rm [NeIII]15.6}}{\rm F_{[NeII]12.8}} \right) - 0.19,
\end{equation}

\noindent where $n_{\rm O^{++}}$ and $n_{\rm O^{+}}$ are the number densities of the O$^{++}$ and O$^+$ ions, respectively, and F$_{\neIII\,15.6}$ and F$_{\neII\,12.8}$ are the observed fluxes in the \neIII\,15.6\um\ and \neII\,12.8\um\ lines, respectively. This relation has a scatter of $\sim$ 0.34 dex. 

\begin{figure*}
\begin{center}	
\begin{tabular}{cc}
\includegraphics[width=0.45\textwidth]{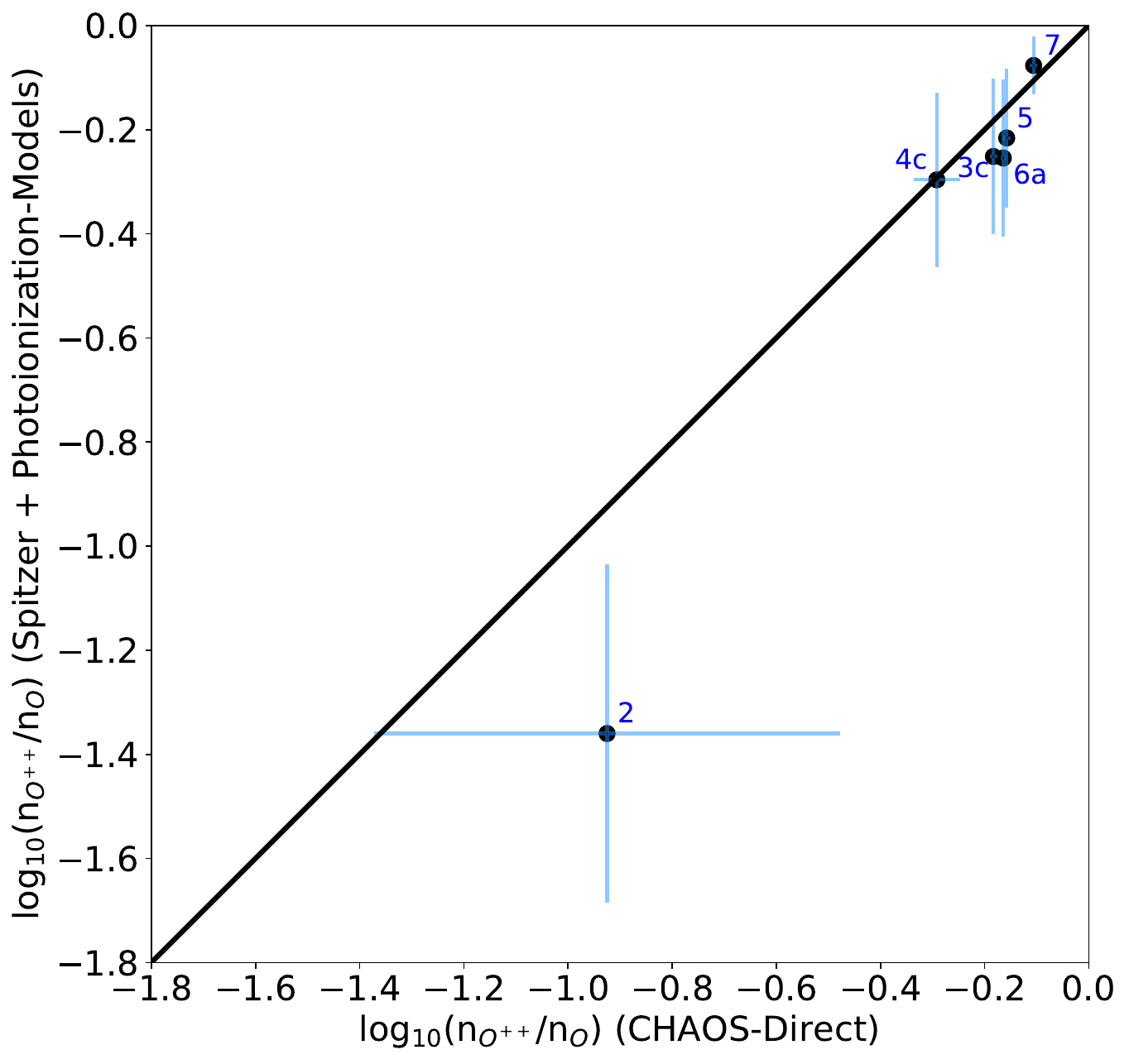}{(a)}
\includegraphics[width=0.45\textwidth]{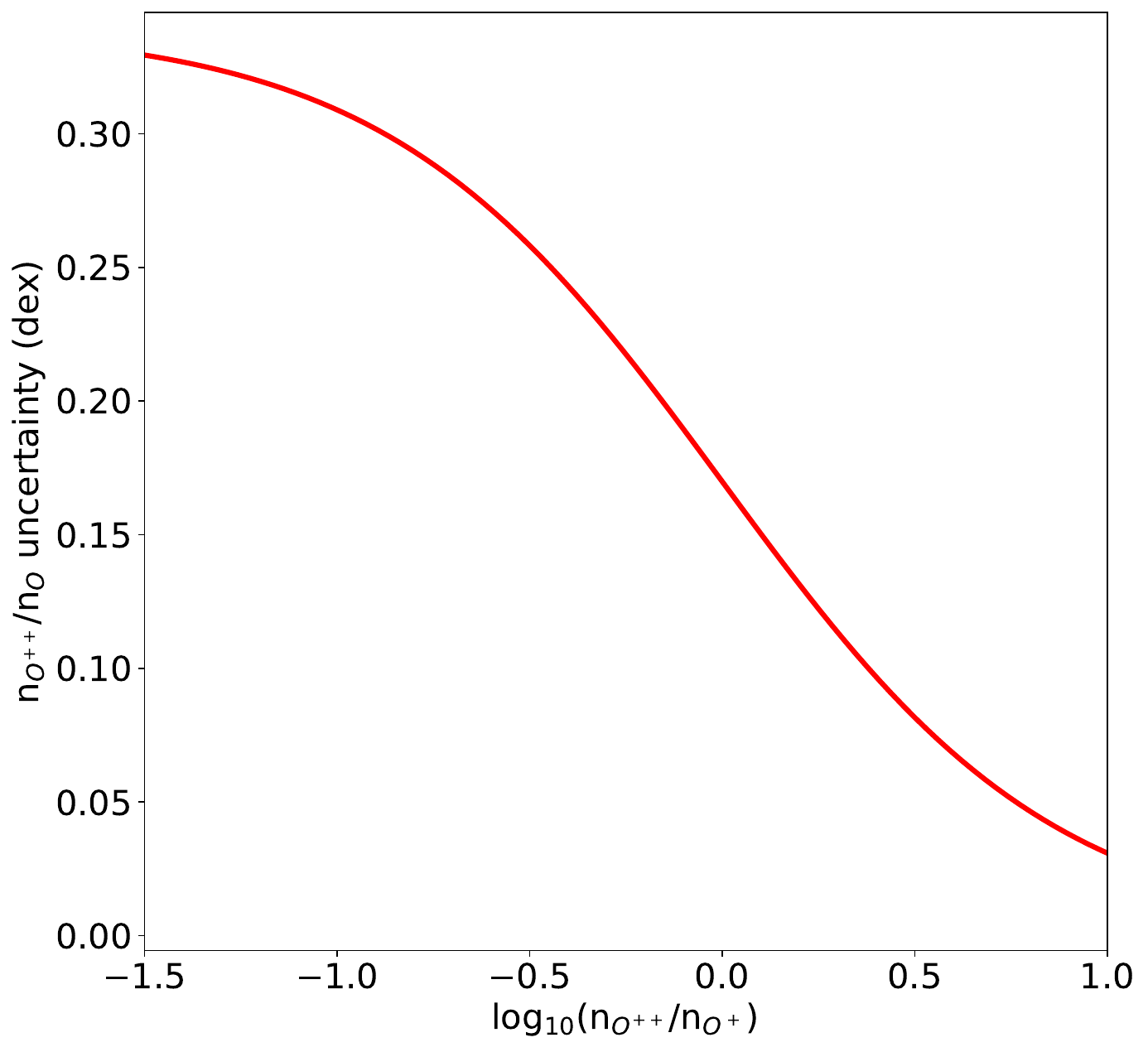}{(b)}
\end{tabular}
\caption{(a) A comparison between the CHAOS-measured and Spitzer+photoionization-model derived ICFs shows that the two ICF calculations yield consistent results in all regions were both CHAOS and Spitzer observations exist. (b) The contribution of the 0.34 dex uncertainty in the \neIII/\neII-derived $n_{\rm O^{++}}$/$n_{\rm O^{+}}$ abundance-ratio from photoionization models to the uncertainty in the derived $n_{\rm O^{++}}$/$n_{\rm O}$ ICF as a function of the $n_{\rm O^{++}}$/$n_{\rm O^{+}}$ abundance-ratio. At large values of the n$_{\rm O^{++}}$/n$_{\rm O^{+}}$ abundance-ratio, where $n_{\rm O^{++}}$/$n_{\rm O}$ $\sim$ 1, uncertainty in the derived $n_{\rm O^{++}}$/$n_{\rm O}$ ICF tends towards zero, as all oxygen is in the doubly-ionized state, but at small values of the $n_{\rm O^{++}}$/$n_{\rm O^{+}}$ abundance-ratio, where $n_{\rm O^{++}}$/$n_{\rm O}$ is small, the 0.34 dex uncertainty in the derived $n_{\rm O^{++}}$/$n_{\rm O^{+}}$ abundance-ratio translates directly to a 0.34 dex uncertainty in the derived $n_{\rm O^{++}}$/$n_{\rm O}$ ICF, and hence absolute oxygen-abundance.}
\label{fig:icfUncertainty}
\end{center}
\end{figure*}

The black points with errorbars in Figure \ref{fig:icf} show the observed n$_{\rm O^{++}}$/n$_{\rm O^{+}}$ abundance-ratio measured by CHAOS, as a function of the \neIII\,15.6\um/\neII\,12.8\um\ line-ratio, observed with Spitzer, in \HII\ regions where both exist. The observed points are in good agreement with the photoionization-model-derived n$_{\rm O^{++}}$/n$_{\rm O^{+}}$ abundance-ratio, with all points lying within the uncertainty of the derived relation (see Figure \ref{fig:icfUncertainty}a).

If we assume that all oxygen within the \HII\ regions is in either the singly- or doubly-ionized state, e.g., $n_{\rm O}$ = $n_{\rm O^{++}}$ + $n_{\rm O^{+}}$, we can relate the n$_{\rm O^{++}}$/n$_{\rm O^{+}}$ abundance-ratio to the $n_{\rm O^{++}}$/$n_{\rm O}$ ratio (ICF) that we need to obtain the total oxygen-abundances:

\begin{equation}
\frac{n_{\rm O^{++}}}{n_{\rm O}} = \frac{1}{1 + \frac{n_{\rm O^{+}}}{n_{\rm O^{++}}}} = \frac{1}{1 + \left( \frac{n_{\rm O^{++}}}{n_{\rm O^{+}}} \right) ^{-1} }.
\end{equation}

The impact of the uncertainty in the ICF decreases as the degree of ionization increases. In Figure \ref{fig:icfUncertainty}b, we have plotted the contribution of the 0.34 dex uncertainty in the derived n$_{\rm O^{++}}$/n$_{\rm O^{+}}$ abundance-ratio from photoionization models to the uncertainty in the derived $n_{\rm O^{++}}$/$n_{\rm O}$ ICF as a function of the n$_{\rm O^{++}}$/n$_{\rm O^{+}}$ abundance-ratio. At large values of the n$_{\rm O^{++}}$/n$_{\rm O^{+}}$ abundance-ratio, where $n_{\rm O^{++}}$/$n_{\rm O}$ $\sim$ 1, uncertainty in the derived $n_{\rm O^{++}}$/$n_{\rm O}$ ICF tends towards zero, as all oxygen is in the doubly-ionized state. At small values of the n$_{\rm O^{++}}$/n$_{\rm O^{+}}$ abundance-ratio, where $n_{\rm O^{++}}$/$n_{\rm O}$ is small, the 0.34 dex uncertainty in the derived n$_{\rm O^{++}}$/n$_{\rm O^{+}}$ abundance-ratio translates directly to a 0.34 dex uncertainty in the derived $n_{\rm O^{++}}$/$n_{\rm O}$ ICF, and hence the absolute oxygen-abundance.

For \HII\ regions in M101, where directly-measured n$_{\rm O^{++}}$/n$_{\rm O^{+}}$ abundance-ratios exist in CHAOS, we employ them to derive the total gas-phase oxygen-abundance from our FIR O$^{++}$ ionic abundances. Figure \ref{fig:icfUncertainty}a shows that the CHAOS- and photoionization-model-derived n$_{\rm O^{++}}$/n$_{\rm O^{+}}$ abundance-ratios are completely consistent, in all \HII\ regions where both are calculated, such that this choice does not affect our results. In this pilot study, only one \HII\ region requires the use of the \neIII\,15.6\um/\neII\,12.8\um\--ICF scaling, where no corresponding n$_{\rm O^{++}}$/n$_{\rm O^{+}}$ ratio is determined by CHAOS. For this region, the nucleus, we employ the derived \neIII\,15.6\um/\neII\,12.8\um\--ICF relations (Eqs. 4 and 5) to obtain the total oxygen abundance (see Table \ref{Absolute_Abundances_Table}).

\section{FIR-Derived Total Abundances}

Applying the ICFs, measured with CHAOS where available, or estimated from the \neIII/\neII\ line-ratio for the single (nuclear) region where CHAOS cannot measure the O$^{++}$ abundance, we convert FIR O$^{++}$ ionic abundances to total gas-phase oxygen abundances:

\begin{equation}
\frac{n_{\rm O}}{n_{\rm H}} = \frac{n_{\rm O^{++}}}{n_{\rm H^+}} \frac{n_{\rm O}}{n_{\rm O^{++}}}.
\end{equation}

\begin{figure*}
\begin{center}		
\begin{tabular}{cc}
\includegraphics[width=0.65\textwidth]{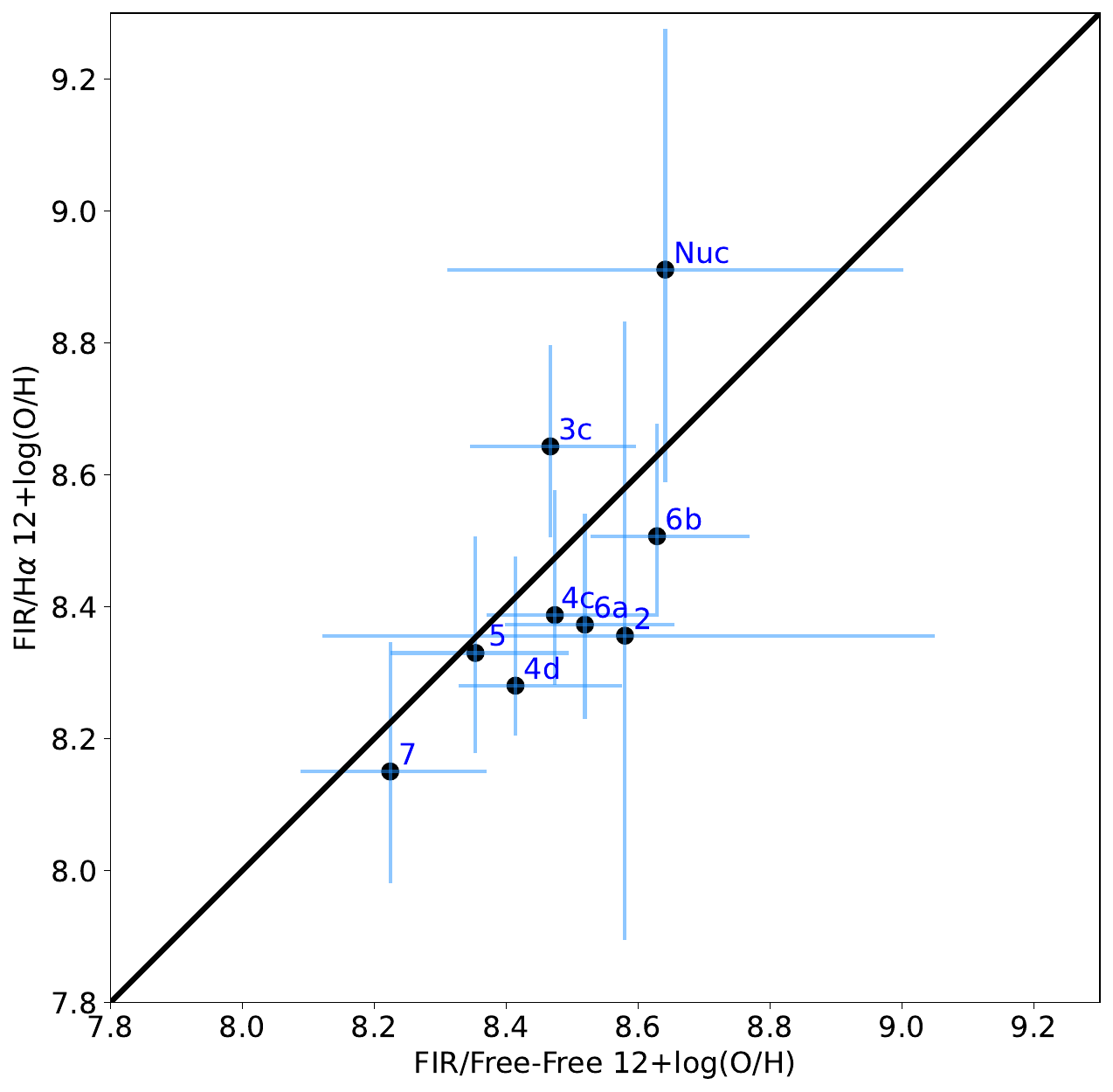}{(a)} \\
\includegraphics[width=0.45\textwidth]{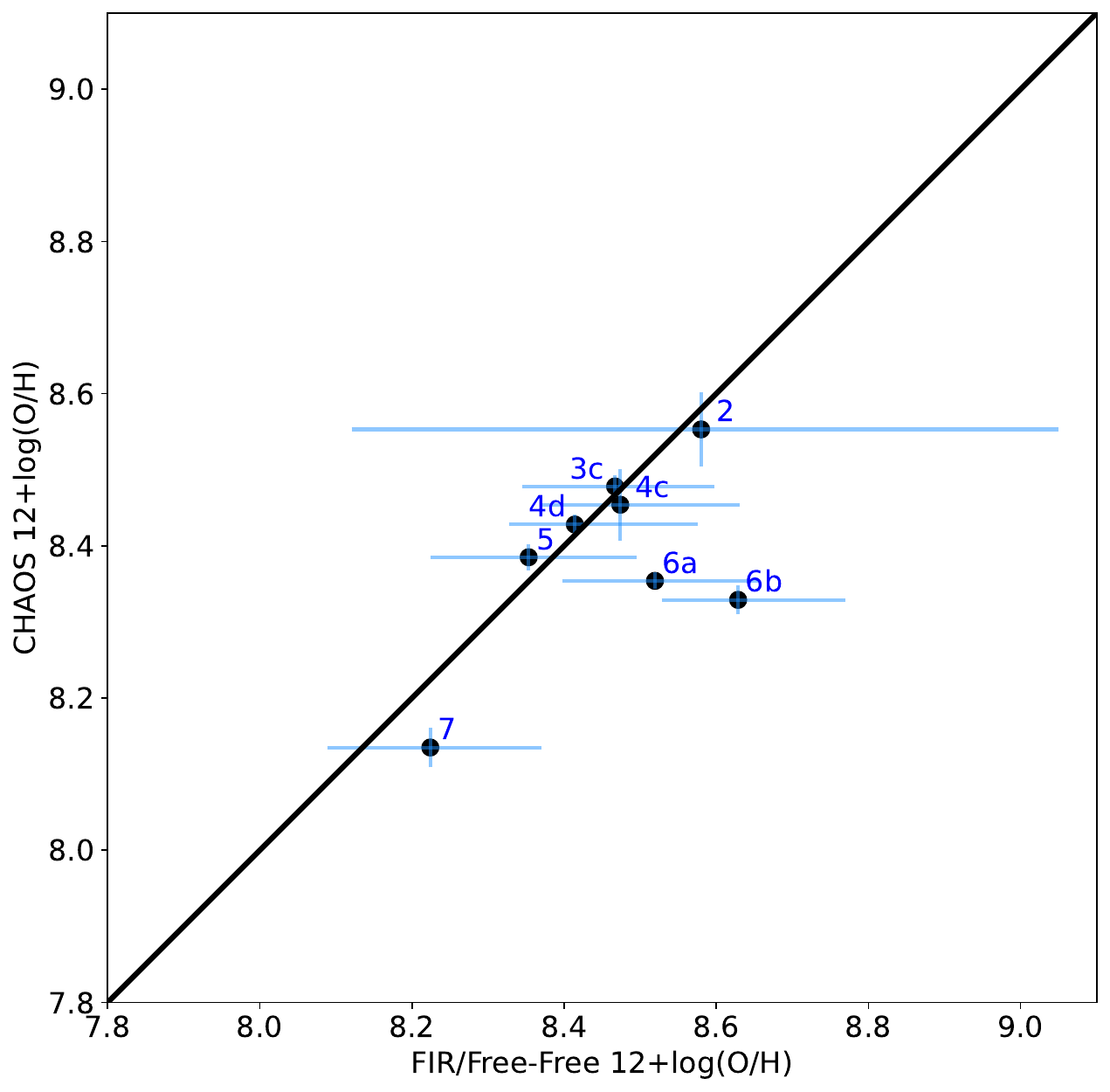}{(b)}
\includegraphics[width=0.45\textwidth]{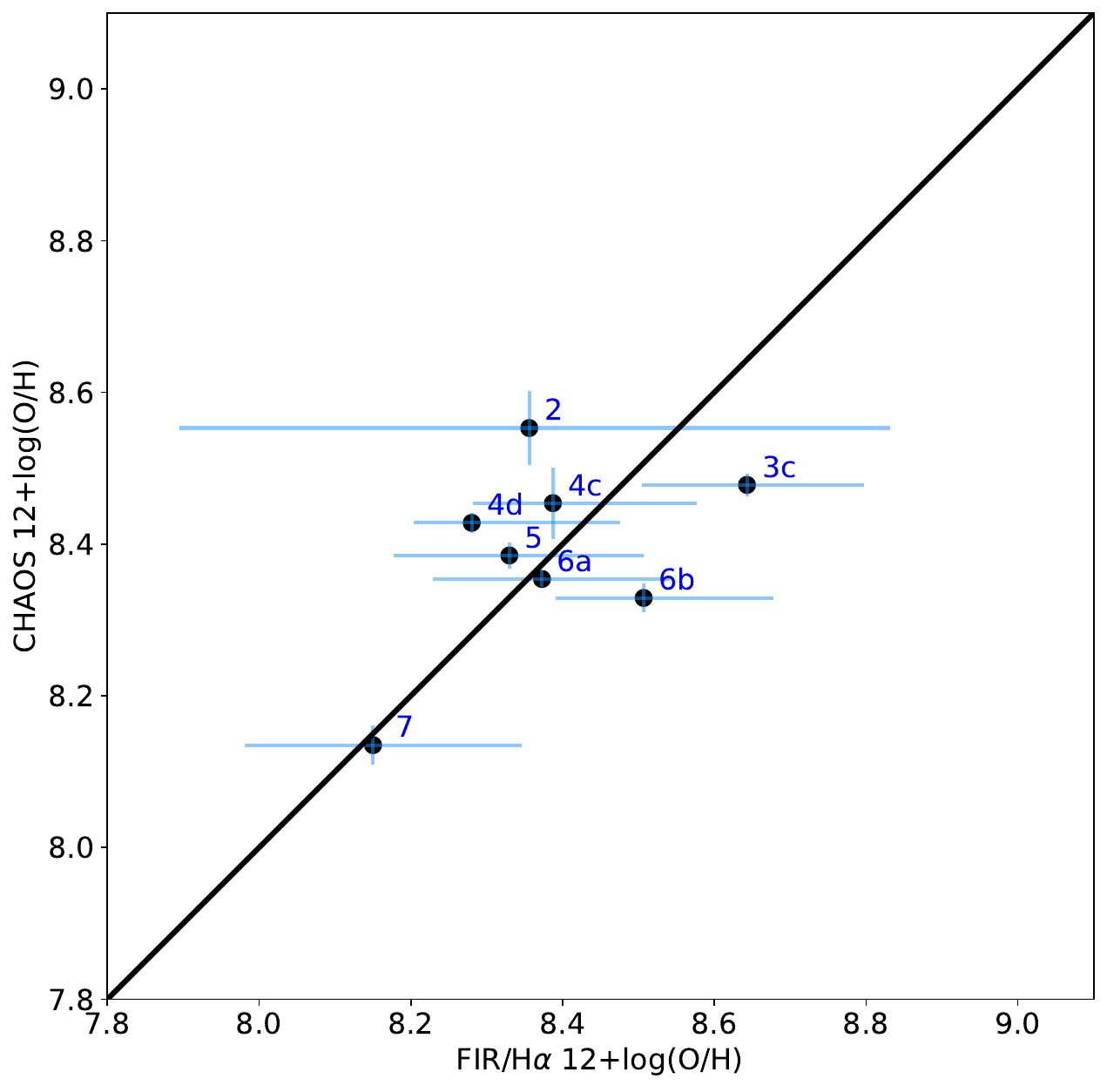}{(c)}
\end{tabular}
\caption{Comparison between the O/H total abundances measured by FIRA (this work) and CHAOS temperature-corrected auroral-line methods \citep[][]{Croxall2016}. (a) O/H abundances derived using the FIR/free-free and FIR/H$\alpha$ normalizations are in excellent agreement, within the scatter of $\sigma_{(12+log(O/H))}$(free-free $-$ H$\alpha$) = 0.15 dex. Similarly, both the FIR/free-free (b) and FIR/H$\alpha$ (c) normalizations are in excellent agreement with CHAOS, within the scatter of $\sigma_{(12+log(O/H))}$(FIRA $-$ CHAOS) = 0.11 dex and 0.13 dex, respectively. \HII\ regions appearing in (a) but absent from (b) and (c) are not detected in the \oIII\,4363\AA\ line by the CHAOS program.}
\label{Absolute_Abundances}
\end{center}
\end{figure*}

Figure \ref{Absolute_Abundances} shows the excellent overall agreement between the direct temperature-corrected abundances from CHAOS and the FIR-derived abundances presented here. As expected from Section 4, we see that the abundances determined using the FIR/free-free hydrogen-normalization are consistent with those calculated using FIR/H$\alpha$ within the standard deviation of 0.15 dex, which is due to the scatter in the observed $F_{\rm H \alpha}$/$S_{\rm ff, 33GHz}$ ratios (see Figure \ref{Absolute_Abundances}a and Table \ref{Absolute_Abundances_Table}). Comparing to the CHAOS-derived abundances, where available, we find that both the FIR/free-free and FIR/H$\alpha$ normalizations produces values that are consistent with the CHAOS direct-abundance method, within the standard-deviations of 0.11 dex and 0.13 dex, respectively (see Figures \ref{Absolute_Abundances}b and \ref{Absolute_Abundances}c).

\begin{figure*}
\begin{center}		
\includegraphics[width=0.95\textwidth]{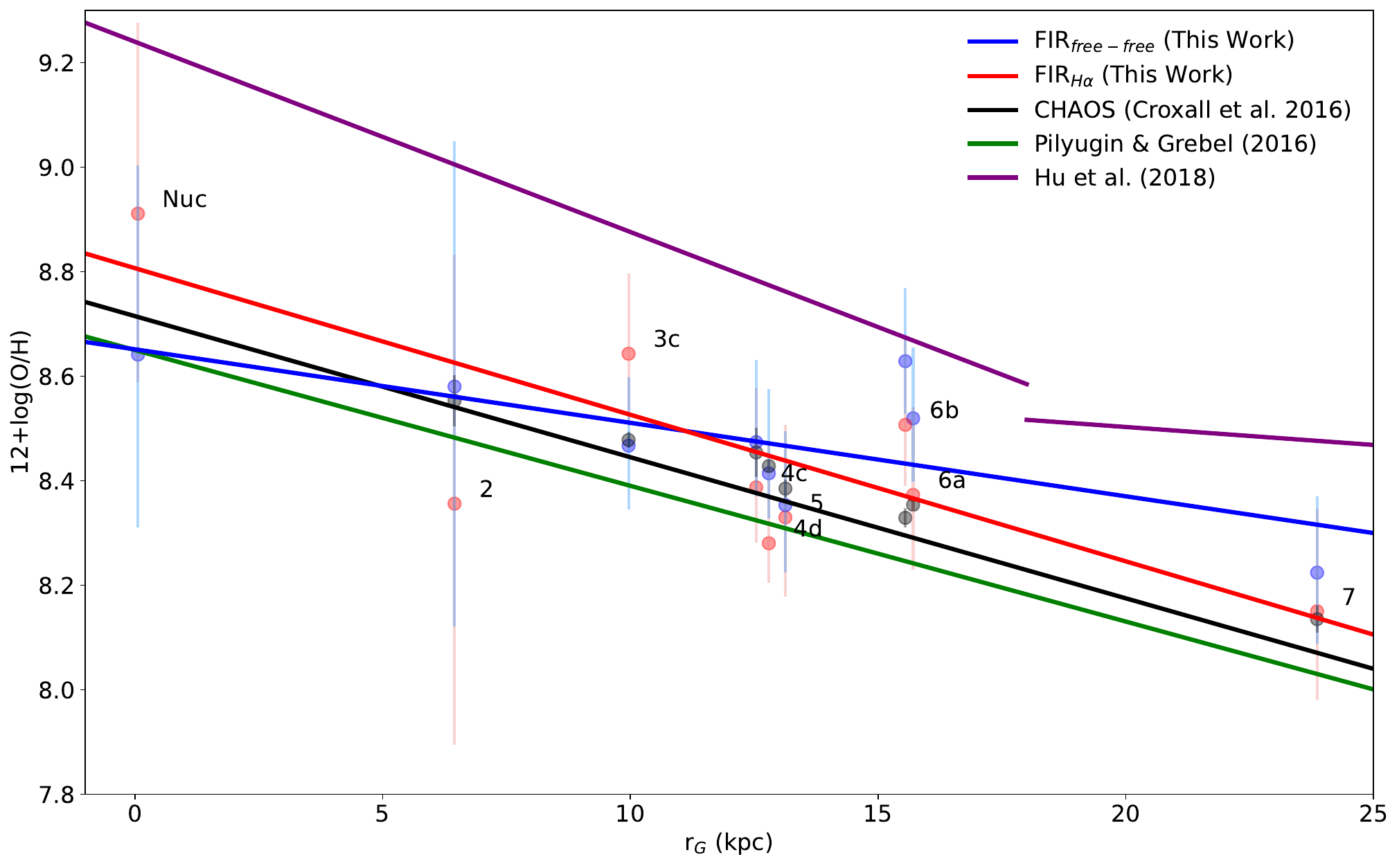}
\caption{Comparison between the total O/H abundance gradient measured by temperature-corrected auroral-line methods \citep[CHAOS; black points, black trend-line fit to all \HII\ regions in the CHAOS dataset,][]{Croxall2016} and the FIRA O/H abundances derived here using both free-free (blue) and H$\alpha$ (red) as the hydrogen normalization. The FIR/H$\alpha$ and CHAOS optical direct-methods are in excellent agreement, while a somewhat flatter abundance-gradient is observed with the FIR/free-free method (at the level of $\sim$ 1.5$\sigma$). The strong-line abundances of \cite{Pilyugin2016} (green line) show similar slope to the CHAOS direct-abundances, though with lower normalization than any of the direct-abundance indicators, while the strong-line calibration of \cite{KK04}, employed in \cite{Hu2018} (purple line), is consistently above the direct-abundance radial-gradients.}
\label{Metallicity_Gradients}
\end{center}
\end{figure*}

We also compare the radial metallicity gradient in M101 obtained from our FIR measurements to those determined from CHAOS direct-abundances (see Figure \ref{Metallicity_Gradients}) as well as strong-line methods calibrated against direct abundances and \HII-region models \citep[e.g.,][]{Pilyugin2016, Hu2018, KK04}. We see a clear radial metallicity-gradient using the FIR-abundance measurements, with a linear fit to the radial gradient using the free-free normalization given by:

\begin{equation}
\begin{split}
12 + \log({\rm O/H})_{ff} = \\
(8.651 \pm 0.124) - (0.014 \pm 0.008)*R_G({\rm kpc}),
\end{split}
\end{equation}

\noindent with a scatter of 0.09 dex, and 

\begin{equation}
\begin{split}
12 + \log({\rm O/H})_{Ha} = \\
(8.807 \pm 0.129) - (0.028 \pm 0.009)*R_G({\rm kpc}),
\end{split}
\end{equation}

\noindent with a scatter of 0.13 dex, for the H$\alpha$ normalization. The slope from the FIR/H$\alpha$ determination, $-$0.028 $\pm$ 0.009 dex\,kpc$^{-1}$, agrees very closely with the CHAOS-derived slope, $-$0.027 $\pm$ 0.001 dex\,kpc$^{-1}$, while the FIR/free-free determination has a flatter slope, $-$0.014 $\pm$ 0.008 dex\,kpc$^{-1}$, in slight tension with the CHAOS-derived value (at the $\sim$ 1.5$\sigma$ level), but still consistent with the FIR/H$\alpha$ slope. We also see that the strong-line calibration \cite{Pilyugin2016} is in reasonable agreement with the direct-abundance radial-gradients, whether FIR or optical, although it falls below them by $\sim$ 0.1 -- 0.15 dex. In contrast, the strong-line calibration of \cite{KK04}, employed in \cite{Hu2018}, lies well above the direct-abundance radial-gradients ($\sim$ 0.4 dex), as is often observed for calibrations based on photoionization models. The reason for the commonly-observed photoionization-model offset from the direct and strong-line methods is not entirely clear, but may include the plane-parallel geometries employed in the modeling, the difficult-to-constrain depletion of metals onto dust grains, or any clumpy structure in the emitting regions \citep[see, e.g.,][and references therein]{KewleyEllison2008}.

\begin{deluxetable*}{cccccc}
\tablecaption{Derived Total Abundances}
\tablecolumns{6}
\tablenum{5}
\tablehead{
\colhead{Region} & \colhead{\neIII/\neII} & \colhead{n$_{\rm O^{++}}$/n$_{\rm O^{+}}$\tablenotemark{a}} & \colhead{O/H (FIR/H$\alpha$)} & \colhead{O/H (FIR/Free-Free)} & \colhead{O/H (Optical Direct-Abundance)\tablenotemark{a}} \\
\colhead{} & \colhead{} & \colhead{} & \colhead{12+log(O/H)} & \colhead{12+log(O/H)} & \colhead{12+log(O/H)} \\
}
\startdata
Nuc & 0.08 \,$\pm$\, 0.01 & 0.073 \,$\pm$\, 0.057\tablenotemark{b} & $ 8.91 ^{+ 0.36 }_{- 0.32 }$ & $ 8.64 ^{+ 0.36 }_{- 0.33 }$ & ... \\
1 & ... & ... & ... & ... & ... \\
2 & 0.05 \,$\pm$\, 0.01 & 0.13 \,$\pm$\, 0.16 & $ 8.36 ^{+ 0.48 }_{- 0.46 }$ & $ 8.58 ^{+ 0.47 }_{- 0.46 }$ & 8.55 \,$\pm$\, 0.05 \\
3a & ... & 0.54 \,$\pm$\, 0.11 & ... & ... & 8.49 +/- 0.03 \\
3b & ... & 0.40 \,$\pm$\, 0.17 & ... & ... & 8.44 +/- 0.05 \\
3c & 2.20 \,$\pm$\, 0.31 & 1.91 \,$\pm$\, 0.09 & $ 8.64 ^{+ 0.15 }_{- 0.14 }$ & $ 8.47 ^{+ 0.13 }_{- 0.12 }$ & 8.48 \,$\pm$\, 0.01 \\
3d & ... & ... & ... & ... & ... \\
4a & ... & ... & ... & ... & ... \\
4b & ... & ... & ... & ... & ... \\
4c & 1.71 \,$\pm$\, 0.24 & 1.04 \,$\pm$\, 0.22 & $ 8.39 ^{+ 0.19 }_{- 0.11 }$ & $ 8.47 ^{+ 0.16 }_{- 0.10 }$ & 8.45 \,$\pm$\, 0.05 \\
4d & ... & 1.74 \,$\pm$\, 0.08 & $ 8.28 ^{+ 0.20 }_{- 0.08 }$ & $ 8.41 ^{+ 0.16 }_{- 0.09 }$ & 8.43 +/- 0.01 \\
5 & 2.77 \,$\pm$\, 0.39 & 2.29 \,$\pm$\, 0.12 & $ 8.33 ^{+ 0.18 }_{- 0.15 }$ & $ 8.35 ^{+ 0.14 }_{- 0.13 }$ & 8.38 \,$\pm$\, 0.02 \\
6a & 2.17 \,$\pm$\, 0.31\tablenotemark{c} & 2.18 \,$\pm$\, 0.07 & $ 8.37 ^{+ 0.17 }_{- 0.14 }$ & $ 8.52 ^{+ 0.14 }_{- 0.12 }$ & 8.35 \,$\pm$\, 0.01 \\
6b & ... & 0.94 \,$\pm$\, 0.08 & $ 8.51 ^{+ 0.17 }_{- 0.12 }$ & $ 8.63 ^{+ 0.14 }_{- 0.10 }$ & 8.33 \,$\pm$\, 0.02 \\
6c & ... & 1.29 \,$\pm$\, 0.06 & ... & ... & 8.42 +/- 0.01 \\
7 & 11.24 \,$\pm$\, 1.59 & 3.63 \,$\pm$\, 0.27 & $ 8.15 ^{+ 0.20 }_{- 0.17 }$ & $ 8.22 ^{+ 0.15 }_{- 0.14 }$ & 8.13 \,$\pm$\, 0.03 \\
\enddata
\tablecomments{The derived absolute O/H abundances, calculated by FIRA using the \oIII\,88\um\ line and either H$\alpha$ or free-free emission for the hydrogen normalization, compared to those calculated by the CHAOS collaboration using collisionally-excited optical direct-abundance methods. \neIII/\neII\ denotes the \neIII\,15.6\um/\neII\,12.8\um\ line-flux ratio (presented here), observed with the SH module of Spitzer/IRS. n$_{\rm O^{++}}$/n$_{\rm O^{+}}$ denotes the abundance ratio of the O$^{++}$ and O$^{+}$ ions.}
\tablenotetext{a}{\cite{Croxall2016}}
\tablenotetext{b}{This n$_{\rm O^{++}}$/n$_{\rm O^{+}}$ abundance-ratio was calculated using the derived \neIII/\neII-ICF scaling (Equations 4 \& 5). It was not measured by CHAOS.}
\tablenotetext{c}{For this region, where a portion of the Spitzer/IRS SH-module field-of-view is located outside of the 20\arcsec\ extraction-aperture, we consider only the portion of the flux contained within the 20\arcsec\ extraction aperture.}
\label{Absolute_Abundances_Table}
\end{deluxetable*}

\section{Discussion}

Understanding the build-up of metals through cosmic time is crucial, as metals significantly affect the physical processes which operate in the ISM. From the formation of stars to the measurement of star-formation rates and molecular-gas mass in galaxies, tracing the build-up of metals through cosmic time is critical to our understanding of galaxy evolution.

Typical methods for determining metal abundances, namely optical strong-line and collisionally-excited direct-methods, suffer from problems in calibration disagreements (in the case of strong-line methods) and exponential temperature-sensitivity (in the case of collisionally-excited optical direct-methods). These optical-based methods must also contend with dust extinction, making them difficult or impossible to employ in the major-mergers of local ULIRGs and vigorously star-forming galaxies at high redshift.

In this context, FIR-based direct-abundance methods are very promising.  The metal-sensitive FIR lines are nearly temperature-insensitive, and can penetrate attenuation from significant dust columns, critical if metallicities are to be accurately measured in local ULIRGs and high-redshift dusty star-forming galaxies. One drawback of these FIR-based methods is in their (departure from quadratic) density dependence and lack of ability to directly measure the ICF in the ionized gas. In this first paper of the FIRA project, we investigated the gas-phase oxygen-abundance in M101 --- critically, a galaxy where excellent overlapping optical direct-abundance determinations exist --- to directly examine the advantages and disadvantages of the FIR-abundance methods.

While examining just one galaxy in this first paper, we obtain measurements that agree exceptionally well with CHAOS collisionally-excited optical direct-abundances, lending strong support for the validity of direct FIR-based absolute-abundances. In particular, we find:

Electron temperatures measured in the O$^{++}$ zone of the \HII\ regions using the \oIII\,5007\AA/88\um\ ratio are broadly consistent with those measured using the optical auroral-line ratio, \oIII\,4363\AA/5007\AA, with an average difference of $\Delta T_e$ = 600 $\pm$ 400 K. This result suggests that any temperature variations, which would affect the two temperature measurements differentially, are small compared to the average temperature uncertainty for each \HII\ region ($\sim$ 1,000\,K). Interestingly, wherever the derived auroral-line and FIR-derived electron-temperatures disagree --- in only two of the \HII\ regions examined here in M101 (each at the \textless\, 2$\sigma$ level) --- the optical auroral-line temperature is higher. We cannot draw any firm conclusions from this observation yet, however it will be interesting to examine whether or not this trend holds in the larger sample. We also find that we can probe \oIII\ temperatures as low as $\sim$ 4,000 K using the \oIII\,5007\AA/88\um\ ratio --- a temperature regime, and hence metallicity regime, that is inaccessible using the \oIII\,4363\AA/5007\AA\ ratio, as the \oIII\,4363\AA\ line becomes impossible to detect at lower excitation, even with highly-sensitive 8m-class spectroscopy. 

While not the focus of the extragalactic FIRA study, the ability to probe such low-temperature gas may be interesting in the context of planetary nebulae as well, since low-temperature gas may be one of the causes of the often-observed abundance discrepancy factors in these objects \citep[e.g.,][]{Liu2006}.

Additionally, we find that the electron densities measured in the \HII\ regions of M101 are always below the critical density of the \oIII\,88\um\ line employed here to determine the gas-phase oxygen-abundance. This finding is consistent with previous measurements, which have shown that much of the ionized gas in the Milky Way and in extragalactic sources is found to be low density \citep[e.g.,][]{Goldsmith2015, Herrera-Camus2016Density}. With the highest-density regions in M101 measured at $n_e \lesssim 300$\,cm$^{-3}$, the effect of the density dependence on the derived oxygen-abundances is only mild, varying by a factor of $\sim$ 2 over the entire range 1 \textless\, n$_e$ \textless\, 300 cm$^{-3}$. We note that the \oIII\,52\um\ line (largely unobserved by Herschel/PACS) would be advantageous to use in place of the \oIII\,88\um\ line for FIR abundances, as it has a much higher critical density \citep[$\sim 3,600$ cm$^{-3}$, as compared to 510 cm$^{-3}$ for the \oIII\,88\um\ line, e.g.,][]{Carilli2013}.

Taken together, the excellent agreement in both the O$^{++}$ zone electron-temperatures measured using the \oIII\,5007\AA/88\um\ and \oIII\,4363\AA/5007\AA\ line-ratios, and in the absolute gas-phase oxygen-abundances derived using the FIRA and CHAOS methodologies, lends strong support for FIR-derived abundance methods. Given that the optical collisionally-excited direct-abundance methods have exponential sensitivity to the electron temperature in the line-emitting regions, while FIR abundances have power-law sensitivity to the electron density in those regions, the agreement between the two methods is quite remarkable. Indeed, the agreement between the FIR and optical abundances, consistent to a level of $\sim$ 0.12 dex, suggests that systematic uncertainties due to temperature fluctuations ($t^2$) are small in the targeted regions of M101, such that the optically-derived abundances are truly representative.  Alternatively, temperature and density fluctuations present within the targeted regions would need to cancel each other out in such a way as to produce consistent abundance values, despite the different underlying parameter dependencies of the temperature- and density-sensitive abundance-methods.

We do see that the scatter in the observed $F_{\rm H \alpha}$/$S_{\rm ff, 33GHz}$ ratios affects the derived abundance-gradients, such that the FIR/free-free abundance-gradient is flatter than is the CHAOS optical direct-abundance gradient (at the level of $\sim$ 1.5$\sigma$). It is difficult to know the cause of the discrepancies in the hydrogen-normalization factors that give rise to these different abundance gradients  -- possibly errors in the applied optical attenuation-correction factors or resolving out radio flux with the VLA interferometer. We will continue to investigate these differences as we expand the FIRA sample to include additional \HII\ regions in other galaxies.

We will also investigate the relationship between the ICF, $n_{\rm O^{++}}$/$n_{\rm O}$, and the derived gas temperature as we expand the FIRA sample. In the M101 data, the bulk of the targeted \HII\ regions have temperatures in the range T $\sim$ 8,000 $-$ 10,000\,K, with just one region at significantly higher temperature (region 7, T $\sim$ 12,000\,K), and a few at lower temperature (the nucleus and region 1, T $\sim$ 4,000 $-$ 5,000\,K). Filling in the missing temperature parameter-space will be critical in determining the relationship between the ICF and gas temperature.

In the near-term future, we will be expanding the FIRA project to include a larger sample of nearby galaxies observed by Herschel/PACS in the \oIII\,88\um\ line. This larger sample will allow us to explore some of the interesting physics hinted at in this single-galaxy paper, including the agreement between auroral- and FIR-derived \oIII\ temperatures, the scatter between the different hydrogen normalizations, and the excellent overall agreement between the optical- and FIR-derived direct oxygen-abundances.

We are also encouraged by the ongoing efforts of SOFIA/FIFI-LS to observe the \oIII\,88\um, 52\um, and \nIII\,57\um\ lines in the local Universe. And in the more distant future, we look forward to observations from a large-aperture space-based FIR observatory, which should revolutionize our understanding of metal abundances in the early universe.

\section{Conclusions}

In this introduction to the far-IR abundances (FIRA) project, we examined the validity of the far-infrared direct-abundance method, comparing it to collisionally-excited optical direct-abundance techniques and strong-line methods, in determining the gas-phase oxygen-abundance in M101. Our main findings are:

\begin{enumerate}

\item O$^{++}$ electron-temperatures measured using the \oIII\,5007\AA/88\um\ line-ratio in 20\arcsec-diameter apertures are consistent with those measured using the \oIII\,4363\AA/5007\AA\ line-ratio in $\sim$ 1\arcsec\ $\times$ 9\arcsec\ slits, with an average difference of $\Delta T_e$ (CHAOS $-$ FIRA) = 600 $\pm$ 400 K.

\item The electron-densities measured in the \HII\ regions of M101 are always below the critical density of the \oIII\,88\um\ line, with the highest-density regions measured at $n_e \lesssim 300$\,cm$^{-3}$. In this moderate- to low-density regime, the effect on the derived abundances is minimal.

\item The measured $F_{\rm H \alpha}$/$S_{\rm ff, 33GHz}$ ratios scatter around the theoretical values for the \HII\ regions of M101, calculated at the derived densities and temperatures for each region individually, with a fractional scatter of $\sim$ 39\%. This scatter may be due to errors in the adopted optical attenuation-correction factors or the radio interferometer resolving out flux on larger scales.

\item The O$^{++}$/H$^+$ ionic abundances derived using the FIR/free-free and FIR/H$\alpha$ normalizations are in good agreement, within the fractional scatter of 39\%, due to the scatter in the hydrogen normalizations. Similarly, both the FIR/free-free and FIR/H$\alpha$ methods are in good agreement with CHAOS optical direct-abundance techniques, within a fractional scatter of 24\% and 29\%, respectively.

\item We derive an indirect O$^{++}$/O ionization-correction-factor (ICF) based on the \neIII/\neII\ mid-IR line-ratio and photoionization models from the 3MdB collaboration. The derived ICF produces \neIII/\neII\ line-flux ratios and $n_{\rm O^{++}}$/$n_{\rm O}$ ICF values that are in excellent agreement with those observed by CHAOS and Spitzer, where available.

\item We derive direct, absolute, gas-phase oxygen-abundances using both FIR/free-free and FIR/H$\alpha$ normalizations, finding excellent agreement between the two, within the standard deviation of $\sigma_{(12+log(O/H))}$(free-free $-$ H$\alpha$) = 0.15 dex. Similarly, both the FIR/free-free and FIR/H$\alpha$ normalizations are in excellent agreement with CHAOS optical direct-abundance techniques, within the standard deviations of $\sigma_{(12+log(O/H))}$(FIRA $-$ CHAOS) = 0.11 dex and 0.13 dex, respectively.

\item We find that the FIR-derived O/H radial abundance-gradient when normalized with H$\alpha$, $-$0.028 $\pm$ 0.009 dex\,kpc$^{-1}$, is consistent with that measured by the CHAOS group, $-$0.027 $\pm$ 0.001 dex\,kpc$^{-1}$, but that both are moderately steeper than the FIR/free-free combination, $-$0.014 $\pm$ 0.008 dex\,kpc$^{-1}$. We further find that the strong-line calibration of \cite{Pilyugin2016} falls slightly below all of the direct-abundance radial-gradients, whether FIR or optical, while the strong-line calibration of \cite{KK04}, employed in \cite{Hu2018}, lies well above the direct-abundance radial-gradients.

\end{enumerate}

\section*{Acknowledgments}

We thank the anonymous referee for the insightful comments and detailed suggestions that helped to improve this manuscript. We also thank Eva Schinnerer and Karin Sandstrom for contributions to acquiring the PPAK data, as well as Bo Peng and Karla Arellano-C\'ordova for helpful discussions that have improved this manuscript. C. Lamarche acknowledges support from NASA ADAP Grant 80NSSC18K0730. K. Kreckel gratefully acknowledges funding from the German Research Foundation (DFG) in the form of an Emmy Noether Research Group (grant No. KR4598/2-1, PI Kreckel).

The National Radio Astronomy Observatory is a facility of the National Science Foundation operated under cooperative agreement by Associated Universities, Inc.

This work is based in part on observations made with the \emph{Spitzer Space Telescope}, which is operated by the Jet Propulsion Laboratory, California Institute of Technology under a contract with NASA.

Based on observations collected at the Centro Astron\'{o}mico Hispano-Alem\'{a}n (CAHA) at Calar Alto, operated jointly by Junta de Andaluc\'{i}a and Consejo Superior de Investigaciones Cient\'{i}ficas (IAA-CSIC).

\software{ppxf \citep{Cappellari2004}, gandalf \citep{Sarzi2006}, Serenity \citep[v2.2.5.1;][]{Sandin2010}, CUBISM \citep{Smith2007}, PopStar \citep{Molla2009}, Cloudy \citep[v17.02;][]{Ferland2017}, HIPE \citep[v15.0.1;][]{Ott2010}}, PAHFIT \citep{Smith2007PAHFIT}, PyNeb \citep{Luridiana2015}

\bibliography{main.bib}

\end{document}